\documentclass[a4paper,fleqn,usenatbib,useAMS]{mnras} 
\usepackage{float}
\usepackage{graphicx}	
\usepackage{amssymb}
\usepackage{amsmath}	

\usepackage{xcolor}

\newcommand{\ergs}{~erg s$^{-1}$}

\newcommand{\nh}{$\rm{N_{H}}$}

\newcommand{\lbol}{$L_{bol}$}

\def\xmm{{\it XMM-Newton}} 
\def\mass{{\it 2MASS}}

\def\chandra{{\it Chandra}}
\def\suzaku{{\it Suzaku}}
\def\nustar{{\it NuSTAR}}

\def\erosita{{\it eROSITA}}

\def\gaia{{\it GAIA}}

\def\swift{{\it Swift}} 
\def\integral{{\it INTEGRAL}} 
 
\def\maxi{{\it MAXI}} 
\def\nicer{{\it NICER}} 
\def\xte{{\it RXTE}}

\title[BCs and optical to X-ray measurements for XRBs]{Average bolometric corrections and optical to X-ray flux measurements as a function of accretion rate for X-ray binaries}
\author[K. Anastasopoulou et al.]{
K. Anastasopoulou,$^{1,2,3,4}$\thanks{E-mail: konstantina.anastasopoulou@inaf.it} A. Zezas,$^{2,3,4}$ J. F. Steiner,$^{4}$ P. Reig$^{2,3}$ \\
$^{1}$ INAF-Osservatorio Astronomico di Brera, Via E. Bianchi 46, I-23807 Merate (LC), Italy\\
$^{2}$ Institute of Astrophysics, Foundation for Research and Technology-Hellas, GR-71110 Heraklion,Greece \\
$^{3}$ Department of Physics, University of Crete, Herakleio, Greece\\
$^{4}$ Harvard-Smithsonian Center for Astrophysics, 60 Garden Street, Cambridge, MA 02138, USA\\
}

\date{Accepted XXX. Received YYY; in original form ZZZ}

\pubyear{2020}

\begin{document}
\label{firstpage}
\pagerange{\pageref{firstpage}--\pageref{lastpage}}
\maketitle

\begin{abstract}
In this paper we use an \xte{} library of spectral models from 10 black-hole and 9 pulsar X-ray binaries, as well as model spectra available in the literature from 13 extra-galactic Ultra-luminous X-ray sources (ULXs).
We compute average bolometric corrections (BC=$\mathrm{L_{band}/L_{bol}}$) for our sample as a function of different accretion rates. We notice the same behaviour between black-hole and pulsars BCs only when ULX pulsars are included. These measurements provide a picture of the energetics of the accretion flow for an X-ray binary based solely on its observed luminosity in a given band. Moreover it can be a powerful tool in X-ray binary population synthesis models. 
Furthermore we calculate the X-ray (2-10 keV) to optical (V-band) flux ratios at different Eddington ratios for the black-hole X-ray binaries in our sample. This provides a metric of the maximum contribution of the disk to the optical emission of a binary system and better constraints on its nature (donor type etc). We find that the optical to X-ray flux ratio shows very little variation as a function of accretion rate, but testing for different disk geometries scenarios we find that the optical contribution of the disk increases as the $p$ value decreases ($T(r)\sim r^{-p}$). 
Moreover observational data are in agreement with a thicker disk scenario ($p<0.65$), which could also possibly explain the lack of observed high-inclination systems.
\end{abstract}


\section{Introduction}
Studies of X-ray binaries (XRBs) are generally conducted in the limited energy pass-bands provided by the available X-ray observatories. The majority of these observatories (e.g. \chandra, \swift, \xmm, \nicer, \maxi) cover the $\sim0.1$ -- $\sim10.0$\,keV band, which encompasses a significant part of the thermal emission of the accretion disk. Therefore these observations have provided important information on the physics of the accretion disks in XRBs, the absorption by local and/or intervening material, and of course their demographics in different environments \citep[e.g.][]{remillard06,fabbiano06,muno09,coleiro13,haberl16,fornasini17}.
However, as demonstrated by X-ray observations extending beyond 10.0\,keV (e.g. with the \xte{}, \integral{}, \nustar{} observatories) the $0.1$--$10.0$\,keV band probes a small fraction of the bolometric output of XRBs, particularly in the intermediate and hard accretion states \citep[e.g.][]{remillard06, done07}. In particular it does not trace the low-energy tail of the accretion disk thermal emission and the hard X-ray tail of the Comptonized hard spectral component.

The bolometric luminosity (\lbol; 0.02-200 keV) is a defining property of XRBs since it provides a direct picture of the energetics of the accretion flow which is inextricably linked to the accretion disk structure and its impact in the surrounding medium \citep[e.g.][]{done07}. The spectral-energy distribution changes significantly with the accretion state of the XRB, presumably in part owing to the associated changes in the geometry of the accretion flow.
Despite its importance, it is not easy to measure the bolometric luminosity of an XRB since that would require simultaneous observations from multiple observatories over a wide wavelength range. We aim to overcome this limitation by determining bolometric corrections (BCs=$\rm{L_{band}/L_{bol}}$) appropriate for XRBs of different compact object types, accretion rates, or accretion states. 
Based on these BCs one could estimate the bolometric luminosity of any XRB based on its observed luminosity in a given band and its accretion rate. 
BCs are also crucial for comparing predictions for the demographics or luminosity distribution of a given X-ray binary population \citep[e.g. from populations synthesis models][]{belczynski08,fragos13a}. These models predict the accretion rate for a given X-ray binary which is then converted to an X-ray luminosity in a given energy band. This conversion depends critically on the spectrum of the accretion flow, which can be parameterized in terms of a bolometric correction.

The characterization of the donor stars of XRBs is a key component for understanding their nature and disentangling the contribution of the accretion disk in the observed optical emission. The X-ray to optical flux ratio in particular has been used for the identification of high-mass XRBs (HMXBs) in nearby galaxies when optical spectroscopic observations are not available \citep[e.g][]{antoniou09}. Similarly, extrapolation of the accretion disk emission to the optical or UV band has been used as a way to estimate its contribution in the measured multi-wavelength emission of their counterparts and the more accurate characterisation of the donor star \citep[e.g.][]{kaaret04, kaaretrev}.
Moreover \citet{vanparadijs94} found that the contribution of the accretion disk to the optical emission of an X-ray binary can be quite important in the case of low-mass XRBs (LMXBs). They studied 18 LMXBs and found that the absolute visual magnitude ($M{_V}$) originating from their disks ranges from 5 to -5, depending on the X-ray luminosity and the outer radius of the accretion disk, affecting significantly the inferred spectral type of donor stars. 

However, distinguishing between accretion disk and donor star optical emission is rather complex.
For example, \citet{kaaret04} found a spectral type of O4V-B3Ib for the ULX counterpart in Holmberg II but this result is true only if there is no significant contribution of optical light from the accretion disk.
For ULX NGC\,1313 X2 \citet{pakull06} measured a B type donor star (using the $\Sigma- M{_V}$ relation of \citet{vanparadijs94}), while 
\citet{liu07} found a spectral type of O7V stating that the contribution of the accretion disk is insignificant, but agreed with \citet{pakull06} in scenarios where they consider the contribution of disk important.

In HMXBs the relative contribution of the disk to the optical emission of the system is expected to be small and mainly originating from the donor star \citep[e.g.][]{vandenheuvel72,treves80}, In low-mass XRBs (LMXBs) the relative contribution from the disk is considered to be the dominant emission process in the soft state,
while in the hard state the coronal emission dominates the X-ray signal, and a small contribution from the donor star and the jet is possible. For example, \citet{russell06} studied quasi-simultaneous optical, near-infrared (NIR), and X-ray observations of 33 XRBs in order to estimate the contributions of various emission processes (disk, donor star, jet), in those sources. They found that for black-hole XRBs (BHXBs) X-ray reprocessing from the disk is the dominant emission process in the optical, while the jet dominates ($\sim$90\%) in the NIR and the hard state. Moreover X-ray reprocessing from the disk dominates the optical and NIR in neutron star XRBs with possible contribution from the jet only at high luminosities.
Therefore a reliable measurement of the optical luminosity (i.e spectral type) of the donor star, unless the XRB is on quiescent, requires the exclusion of any contamination from the accretion disk. Since observationally this is very challenging, a theoretical calculation of the fraction of optical to X-ray flux ($\rm{\xi = f_{opt}/f_{x}}$)  in a given band will make possible the calculation of the optical flux of the accretion disk only with the knowledge of its X-ray flux which could be then used in any study of the optical emission of the XRBs.

In this paper we use $\sim$10000 model spectra of young XRBs (pulsars; black-holes; $<$100\,Myr) and create prescriptions for calculating average BCs as well as the average $\xi$ values for the accretion disk as a function of the accretion rate. In Section \ref{dataandanalysis} we present the data used in this study as well as the analysis procedure. We discuss our results and their implications in other studies in Section \ref{discussion}. Finally in Section \ref{summary} we summarise our findings.

\section{Data}\label{dataandanalysis}

This paper focuses on the analysis of X-ray spectral models of 10 Galactic BHXBs, 9 pulsar XRBs and 13 extra-galactic ultra-luminous X-ray sources (ULXs). The scope is to understand the average emission properties (BCs, $\xi$) of XRBs at different accretion rates, and for that reason we selected XRBs with reliable distance and mass measurements. Detailed information of known physical properties of the selected sample of BH, pulsar XRBs, and ULXs and their corresponding references are reported in Tables \ref{tab.bhs}, \ref{tab.pulsars}, and \ref{tab.ulxs} respectively. For pulsar XRBs as well as for confirmed ULX pulsars (ULXps), we assume a mass of 1.4 M$_{\odot}$. For the rest of the ULXs we either assume that they are all neutron stars with a mass of 1.4 M$_{\odot}$, or BHs with a mass of $9.1\,M_{\odot}$, which is the average mass of the Galactic BHs included in our sample (see Table \ref{tab.bhs}).

\subsection{The spectral models}\label{spectralmodels}

\subsubsection{Black-hole X-ray binaries}\label{BHs}
In this work we used the BH XRB spectral models (Table \ref{tab.bhs}), for 10 out of the 29 stellar-mass BH and BH candidates (total 15,000 spectra), reported in \citet{steiner16}, and collected over the 16 years of Rossi X-ray Timing Explorer (RXTE) operations. 
The selection of the 10 systems was based on the existence of reliable distance and mass measurements. The spectra were fitted with a simplistic spectral model of 
\texttt{phabs$\times$[smedge(simpl$\otimes$ diskbb)+Gauss]} (hereafter Model 1), where 
\texttt{phabs} represents the absorption along the line of sight, \texttt{diskbb} and \texttt{simpl} the disk and the Compton components respectively, \texttt{smedge} the relativistically broadened Fe–K absorption edge, and \texttt{Gauss} the reflection generated by illumination of the cold disk by the power-law component” which is used to fit a relativistically smeared Fe K$\alpha$ line, or in some cases the absorption due to an outflow.
Moreover, we used three other spectral models just for comparison purposes in our analysis. These spectral models are \texttt{phabs$\times$[smedge(simplcut$\otimes$ diskbb)+Gauss]} (hereafter Model 2), where \texttt{simplcut} is a Compton component with a cut off at high energies, \texttt{phabs$\times$[smedge(simpl$\otimes$ diskbb)+laor]} (hereafter Model 3), where \texttt{laor} represents a relativistically broadened Fe-line fluoresced from the disk by coronal illumination and 
\texttt{phabs$\times$[(simpl$\otimes$ diskbb)+Gauss]} (hereafter Model 4).

\subsubsection{Pulsar X-ray binaries}

Accreting pulsars is a dominant component of the HMXB populations \citep[e.g.][]{reig11}. 
In particular, HMXBs with a main-sequence companion -- the so-called Be/X-ray binaries-- represent the most numerous group of X-ray pulsars \citep{reig11,paul11}. They are hard X-ray transient systems that remain most of their lives in quiescence. Occasionally they undergo giant X-ray outbursts. When this occurs, the X-ray luminosity changes by 3 or 4 orders of magnitude, reaching Eddington values. Because of these large amplitude variations in luminosity, BeXBs represent excellent laboratories to test accretion models. In this work, we analysed 9 accreting pulsars. These are all the systems that have been through a giant outburst (type II) during the RXTE life time, and also have reliable distance measurements. The data were retrieved from the {\it RXTE} archive and covered an X-ray outburst. The broadband spectra were fitted with a simple phenomenological model consisting of an absorbed power-law model modified at high energies by an exponential cutoff,
\texttt{phabs$\times$cutoffpl}. The discrete components (spectral lines) to account for Fe K fluorescence and cyclotron resonant scattering features when present were fitted with a narrow Gaussian (\texttt{Gauss}), and Lorentzian (\texttt{cyclabs}) profiles, respectively. The continuum fitted with a power-law leaves residuals in the 10--20 keV energy range in most cases. We added a broad Gaussian in emission to account for these residuals \citep[see e.g.][]{klochkov08}.

\begin{table*}
 \centering
		\begin{minipage}{140mm}
		\caption{Black-hole X-ray binaries}
		\begin{tabular}{@{}lcccccc@{}}
			\hline 
			Name& Distance & Reference& Mass & Reference& Spectral Type &$\mathrm{N_{models}}$ \\
			& kpc& &$M_{\odot}$ & & &\\
			\hline
			XTE J1118+480 & $4.16_{-2.12}^{+3.91}$ & G19 &6.9-8.2& K13a & K7-M1V & 124\\[2pt]
			GRO J1655-40 & $3.74\pm0.50$& G19 &$6.6\pm0.5$ & SH03 & F6IV& 987 \\[2pt]
			XTE J1550-564	&$4.40\pm0.50$& O02, O11 &$9.1\pm0.6$& O11& K3III& 517\\[2pt]
			GRS 1915+105 &$8.60\pm2.00$& G19 &$12.5\pm2.0$ & R14 & K1-5III& 2566\\[2pt]
			V4641 Sgr	&$6.24_{-1.34}^{+2.04}$& G19 &$6.4\pm0.6$& M14& B9III& 94\\[2pt]	
			4U 1543-475		&$8.60_{-2.46}^{+3.84}$& G19 &$8.4-10.4$& O98, O03& A2V& 130\\[2pt]
			MAXI J1659-152	&$8.60\pm3.70$& K13b &$6.0-8.0$\footnote{for a distance of 8.6 kpc} & Y12 & M5V& 71\\[2pt]	
			Cyg X-1&$2.38_{-0.17}^{+0.20}$& G19 &$14.8\pm1.0$& T14 & 09.7Iab& 2446\\[2pt]
			LMC X-1 &$48.10\pm2.20$& S14&$10.9\pm 1.5$ & O09 & O8 & 704\\[2pt]
			LMC X-3&$48.10\pm2.20$&S14&$7.0\pm 0.6$& O14& B2.5V& 1598\\	[2pt]					
			\hline			
			\end{tabular} 	
		\smallskip
			
		{\textbf{References.} G19---\citet{gandhi19}, K13a--\citet{khargharia13}, K13b---\citet{kuulkers13}, M14---\citet{macdonald14} ,O98---\citet{orosz98}, O02---\citet{orosz02}, O03---\citet{orosz03}, O09---\citet{orosz09}, O11---\citet{orosz11}, O14---\citet{orosz14}, 
		R14---\citet{reid14} SH03---\citet{shahbaz03},
		S14---\citet{steiner14},
		T14---\citet{tomsick14}, Y12---\citet{yamaoka12}}, Spectral types have been adopted from \textit{BlackCAT} catalogue \citep{blackcat} and references therein.
	\label{tab.bhs}
	\end{minipage}
	\end{table*}

\begin{table*}
	\centering
		\begin{minipage}{140mm}
		\caption{Pulsar X-ray binaries}
		\begin{tabular}{@{}lcccccc@{}}
			\hline 
			Name& Distance & Spectral type&Reference &$\mathrm{N_{models}}$ & $E(B-V)$ & \nh \\
			& kpc& & && & $10^{22}\ cm^{-2}$ \\
			\hline			
		   4U 0115+63    &  7.19$_{-1.09}^{+1.47}$ &B0.2Ve & PF15&26 & 1.33& 0.91\\[2pt]
		  1A 0535+262    & 2.12$_{-0.20}^{+0.25}$ &O9.7IIIe  &PF15 &31& 0.74 & 0.51 \\[2pt]
		   1A 1118-616   &  2.92$_{-0.21}^{+0.25}$ &O9.5III-Ve &R11 & 32 & 0.92 (R11)& 0.63\\[2pt]
		  EXO 2030+375   & 3.63$_{-0.88}^{+1.34}$ &	B0Ve & PF15& 133 & 1.65& 1.13\\[2pt]
		  GX 304-1     & 2.01$_{-0.13}^{+0.14}$ & B2Vne &R11 & 21 & 2.0 (R11) & 1.37\\[2pt]
		   KS 1947+300   &  15.19$_{-2.73}^{+3.66}$ & B0Ve &PF15&115 &0.90 & 0.62\\[2pt]
		  MXB 0656-072   &  
		  5.11$_{-0.93}^{+1.35}$ &O9.7Ve & B09 &74 & 0.812& 0.56\\[2pt]
		   V0332+53    &   5.13$_{-0.75}^{+1.01}$ & O8.5Ve & PF15 &80 &1.64& 1.12\\[2pt]
		  XTE J1946+274   & 12.64$_{-2.87}^{+3.88}$ & B0-1IV-Ve & PF15 &65 & 1.65& 1.13\\[2pt]		  	
	\hline			
			\end{tabular} 	
		\label{tab.pulsars}
		\smallskip
	
		\textbf{References.} B09---\cite{belczynski09}, R11---\citet{reig11}, RF15---\citet{reig15} , for the distances \gaia{} measurements were used from \cite{bj18}.
	\end{minipage}
	\end{table*}

\begin{table*}
	\centering
		\begin{minipage}{140mm}
		\caption{Ultra-luminous X-ray sources}
		\begin{tabular}{@{}lccccc@{}}
			\hline  
			Name& Distance & References &Accretor&$\mathrm{N_{models}}$ & References\\
			& Mpc& &&& \\
			\hline
			NGC\,7793 P13 & 3.50& P10 & Pulsar& 1 & W18A \\[2pt]
 			NGC\,300 ULX1 & 1.88& G05 & Pulsar & 1& C18  \\[2pt]
			NGC\,5907 ULX1 & 17.06& T16 & Pulsar & 2& W15, F17 \\[2pt]
			M82 X-1 & 3.30& F14 &Unknown& 4&   B20  \\[2pt]
			IC\,342 X-1 & 3.40& T10 &	 Unknown& 3& R15\\[2pt]
			IC\,342 X-2 &3.40& T10 &		 Unknown& 1&R15, S17\\[2pt]
			NGC\,1313 X-1 & 4.20& M2, T16 & Unknown& 8& W19\\[2pt]
			 NGC\,5204 X-1 & 4.90& T16 &	 Unknown& 1&M15\\[2pt]
			 Circinus ULX5& 4.20& F77 & Unknown& 1&W13\\[2pt]
			Holmberg IX X-1 & 3.55& P02, T16 & Unknown&6& W17 \\[2pt]
			 Holmberg II X-1 & 3.40& K02, T16 &Unknown& 1&W18B \\[2pt]
			 	M33 X-8		 & 0.83& B16 & Unknown&1&WE19\\[2pt]
			NGC\,5643 X-1 &13.90& S03 & Unknown& 2&K16\\
			
	\hline			
			\end{tabular} 	
		\label{tab.ulxs}
		\smallskip
		
		{\textbf{References.} B16---\citet{bhardwaj16}, B20---\citet{brightman20}, C18---\citet{carpano18}, G05---\citet{gieren05}, F77---\citet{freeman77}, F14---\citet{foley14}, F17---\citet{fuerst17}, K02---\citet{karac02}, K16---\citet{krivonos16}, M2---\citet{mendez02}, M15---\citet{mukherjee15},
	P02---\citet{paturel02}, P10---\citet{pietr10}, R15---\citet{rana15}, S03---\citet{sanders03}, S17---\citet{shidatsu17}, T10---\citet{tikhonov10}, T16---\citet{tully16}, W13---\citet{walton13}, 	  
	W15---\citet{walton15},	W17---\citet{walton17}, W18A---\citet{walton18a},
	W18B \citet{walton18b},	W19---\citet{walton19}, 
	 WE19---\citet{west19}	 }
	\end{minipage}
	\end{table*}

\subsubsection{Ultra-luminous X-ray sources}
In order to include in our analysis sources at very high accretion rates we used all ULX spectral models, for which details on their best-fit parameters, were available from the literature (see ``references" column of Table \ref{tab.ulxs}). These models are usually the result of simultaneous spectral fitting from \textit{NuSTAR} data, which extend to harder energies ($>$ 10 keV), and \xmm{}, \chandra{} or \suzaku{} data, which cover softer energies (up to $\sim$10\,keV). For the selection of the appropriate models for this work, we used wherever it was possible physical models that describe the non-thermal component of the spectra, like Compton scattering of seed photons from the disk etc. Those models also describe well the soft part of the unabsorbed model (i.e show a convergence at soft energies; \texttt{comptt}, \texttt{simpl}).
On the contrary the power-law component which is widely used to describe non-thermal emission from XRBs is a good phenomenological model to describe the spectra of XRBs, however it diverges unphysically at energies well below the thermal peak \citep[e.g.][]{steiner09}.

\subsection{The hydrogen column density}\label{nh}

Throughout this work, and for each source we have adopted different values for the hydrogen column density (\nh) in our spectral models. The values used, depend on the class of the source (BHXRB, pulsar XRB or ULX) and the limitations introduced by the spectral modelling.
We explain in more detail below, the reason behind this approach and the different values of the \nh\ values adopted:
\begin{enumerate}

\item The original \nh\ which is the \nh \ resulting from the spectral fitting. For the BHs this value was fixed at a specific value for the same object (see Table 1 of \citet{steiner16}) while for the pulsar XRBs the value was a free parameter during the fitting process. For the ULXs, it was the value provided in the corresponding papers (see References Table \ref{tab.ulxs}).

\item A fiducial value of $0.1 \times 10^{22}\ atoms\ cm^{-2}$, for a Galactic source which is relatively unobscured,  was adopted in order to have a uniform absorption for all models and spectra, and for easier comparison.

\item The \nh\ was set to the value of 0, in order to get the intrinsic spectral shape. In this way, we calculated the physical properties of the sources without the absorption hindering their luminosity and altering the properties of the spectrum especially at soft energies ($<$2\,keV). 
Since the spectral models for the pulsar XRBs are based on power-law models, the estimated unabsorbed flux diverges as the energy band extends to lower energies. Therefore it was not possible to calculate the unabsorbed flux by setting the \nh{} to 0.
This was not possible as well, for 7 out of 13 ULXs (NGC 300 ULX1, NGC 5907 ULX1, Holmberg II X-1, IC 342 X-1, IC 342 X-2, M33 X-8, and NGC 5643 X-1 ), where phenomenological models (e.g. power-law etc.) have been used.

\item 
In order to avoid the divergence of a power-law model, we have assumed for pulsar XRBs absorption due to the circumstellar material with a line-of-sign extinction calculated through the \texttt{Bayestar3} model \citep{Green19} which provides a 3D dust reddening map based on Gaia distances and photometric data from \textit{Pan-STARRS} 1 and \mass{}. We assumed an $\rm{R_{V}=A_{V}/E(B-V)=3.1}$ reddening law to derive the extinction $\rm{A_{V}}$ from the reddening, and the \citet{extinction} relation to calculate the equivalent line-of-sight \nh\ column density from the extinction. 
It was not possible for the reddening to be calculated for sources 1A~1118-616 and GX~304-1 due to the lack of measurements from the \texttt{Bayestar3} model in the vicinity of these sources, and therefore the reddening values reported in \citet{reig11} were used instead.
The extinction, as well as the intrinsic \nh{} are reported in the last two columns of Table \ref{tab.pulsars}.
\end{enumerate}

\subsection{Energy bands}
The energy bands we present along this paper are the bands reported in Table \ref{tab.bands}. They are the ``standard" bands of most X-ray observatories like \chandra, \xmm, \erosita, \nustar. We also use a bolometric band which we define at $\mathrm{0.02-200\,keV}$.

\begin{table}
	\caption{Energy bands}
		\begin{tabular}{@{}ccc@{}}
			\hline  
			&Name & Band (keV) \\
			\hline
			\hline
			Bolometric &Bol & 0.02-200\\			
			\hline	
			Broad&B1 & 0.5-10\\
			&B2 & 4-25\\
			\hline			
		Soft&S1& 0.5-2\\
			&S2& 4-12\\
			\hline
			&H1 &2-10\\
		Hard	&H2& 12-25 \\			
			\hline			
			\end{tabular} 	
		\label{tab.bands}
		\end{table}

\subsection{Analysis}

In the following section we present the basic analysis performed in order to further process the model spectra, which are described in Section \ref{spectralmodels}. This was done for the purposes of calculating BCs, and the contribution of the disk to the optical flux of the binary system. 

\subsubsection{Calculations of Bolometric Corrections}\label{BCs}

In order to calculate the bolometric luminosity (band Bol) for all our sources, we have used the XSPEC v12.11.1 software \citep{xspec}.
In more detail, we extrapolated the spectra using the command \textit{energies} of XSPEC to cover the energy range 0.02-200 keV, which corresponds to the band of the bolometric luminosity. This extrapolation is being made under the assumption that the physical mechanisms that produce the shape of the spectrum we observe, remain unchanged, and no new mechanism is introduced at softer or harder energies. However, caution is advised when using the pulsar XRBs BCs, since a soft excess (modelled usually as a black-body with $kT_{BB}\sim 0.1$ keV) is typically present in the spectra of pulsar XRBs. This soft excess is thought to originate from reprocessing of hard X-rays from the neutron star by the inner region of the accretion disk \citep[][]{hickox04}.

As mentioned above, in order to calculate BCs, necessary was the measurement of the bolometric luminosity for each one of the sources in our sample,  and therefore the extrapolation of all spectral models in order to cover the 0.02 to 200 keV energy range. This method was performed for the absorbed and fiducial absorption model spectra for all classes of sources.
For the calculation of BCs free of the Galactic absorption introduced by each source, the unabsorbed BH spectral models, as well as, the intrinsic \nh{} pulsar XRB spectral models were also extrapolated to cover the bolometric energy range.

Then, we calculated the fluxes for all spectral models, using XSPEC command \textit{flux} for all energy bands provided in Table \ref{tab.bands}. Fluxes were converted to luminosities, assuming spherical symmetry, and using the ``distance" columns in Tables \ref{tab.bhs}, \ref{tab.pulsars}, and \ref{tab.ulxs}.

The Eddington luminosity of each source was calculated using the following:
\begin{equation}
\rm{L_{edd}}=1.3\times 10^{38}\frac{M}{M_{\odot}}ergs^{-1}
\end{equation}
The Eddington ratio, which provides a metric of the accretion rate of material originating form the disk onto the compact object, is defined as:
\begin{equation}
Eddratio=\frac{L_{bol}}{L_{edd}}
\end{equation}
and was calculated for each model spectrum in our sample.
Black hole systems were aggregated into 10 Eddington ratio bins and pulsar XRBs into 5 Eddington ratio bins,
spaced logarithmically starting from the lowest to the highest Eddington ratios. Each bin contains at least 15 spectra, and their range in Eddington ratios can be seen in the second and third Columns of Tables \ref{tab.bcbhs}, \ref{tab.bcpulsars}, and \ref{tab.bcpulsars_in}.

We then computed the BCs for each spectral model as:
\begin{equation}
BC=\frac{L_{band}}{L_{bol}}
\end{equation}
where ``band" is one of the energy bands specified in Table \ref{tab.bands}.

Finally for each Eddington ratio bin we calculated the median value (50 percent percentile) of the BCs, as well as their 1$\sigma$ errors (68 percent percentiles).
We present our results for the average bolometric correction at different accretion rates in Figs. \ref{fig.bc_bh_model1} and  \ref{fig.bh_bc}, and Tables \ref{tab.bcbhs} to \ref{tab.bcbhs4}, for BHs and in Figs. \ref{fig.pulsars_bc_intrisic} and \ref{fig.pulsars_ulxs}, and Tables \ref{tab.bcpulsars}, and \ref{tab.bcpulsars_in} for the pulsar XRBs and ULXs. Examining the behaviour of the BCs, for example at band 0.5-10 keV, and for the different classes of XRBs, we notice that for BH systems (Fig. \ref{fig.bc_bh_model1}) the higher values of BCs are present at lower and higher Eddington ratios, revealing that the bulk of the bolometric X-ray emission in those ratios, is emitted at the energy range of 0.5-10 keV. Regarding the pulsar XRBs and the 0.5-10 keV band, higher BC values are present only at lower Eddington ratios (Fig. \ref{fig.pulsars_bc_intrisic}), but when in our sample we include the ULXs (Fig. \ref{fig.pulsars_ulxs}) we observe the higher BC values at higher Eddington ratios, which is the same pattern observed in BH systems. We discuss our results in more detail and for different energy bands in Section \ref{discussion}.

\subsubsection{Calculation of disk's optical contribution}\label{optical}

In this section we present the analysis performed for the calculation of the XRB optical luminosity originating from the accretion disk and the comparison with its X-ray luminosity at different accretion rates.
This test can only work under the assumption that the same physical mechanisms responsible for the observed spectrum in the X-rays are responsible also for the emission of the disk in the optical.
Therefore we use only the BH XRB unabsorbed model spectra (Table \ref{tab.bhs}) since these spectra are described by physical models (\texttt{simpl}) and could in principle be extrapolated down to lower energies and the optical regime. In addition none of these models contains jet emission or stellar companion emission, and therefore are representative of the disk/corona emission from the system. In our analysis we consider only Model 1 out of the four available models for BHXBs, since no significant difference between the different models (see discussion in Section \ref{BC_BHs}) was observed.

In order to measure the optical flux, we extrapolated the model spectra, using the command \texttt{energies} of XSPEC, down to $10^{-4}$ keV. We then measured the optical flux of the disk using the tool PYPHOT\footnote{\url{https://mfouesneau.github.io/docs/pyphot/}}. The tool computes the photometry of input stellar spectra using a library that contains a significant amount of common filters. We calculated the flux in the V band, using the filter "GROUND\_JOHNSON\_V" from the PYPHOT filter library, with effective wavelength 5438.689 \AA.

We then calculated, the ratio of the optical luminosity $L_V$ and the X-ray luminosity in the 2-10 keV band ($L_X$), at different accretions rates. 
Throughout the text often we refer for simplicity to the X-ray luminosity originating from the disk, corona and reflection components as "disk X-ray luminosity".
The methodology was the same as presented in Section \ref{BCs}, but this time we used L(2-10 keV)/$L_{edd}$ instead of $\mathrm{L_{bol}/L_{edd}}$. We selected the 2-10 keV energy band, because is one of the bands most commonly used in X-ray observatories and in studies of individual XRBs, and therefore our results can be compared easier with observations.
In Table \ref{tab.fxfopt} we present the values of the optical to X-ray emission from the disk, for 10 bins at different Eddington ratios ($\mathrm{L_X/L_{edd}}$), and in Fig. \ref{fig.lxlopt} we show how the ratio of optical to X-ray luminosity varies as a function of different X-ray to $L_{edd}$ ratios. We discuss our results in more detail in Section \ref{optical_discussion}.

\section{Results and discussion}\label{discussion}

\subsection{Bolometric corrections}

In this section we discuss the results of the average BCs for BH and pulsar XRBs based on the unabsorbed model spectra for the BHs and the fiducial and intrinsic absorption for pulsars and ULXs. 

\subsubsection{Black-hole XRBs}\label{BC_BHs}

\begin{figure}
	\centering
		\includegraphics[width=1.0\columnwidth]{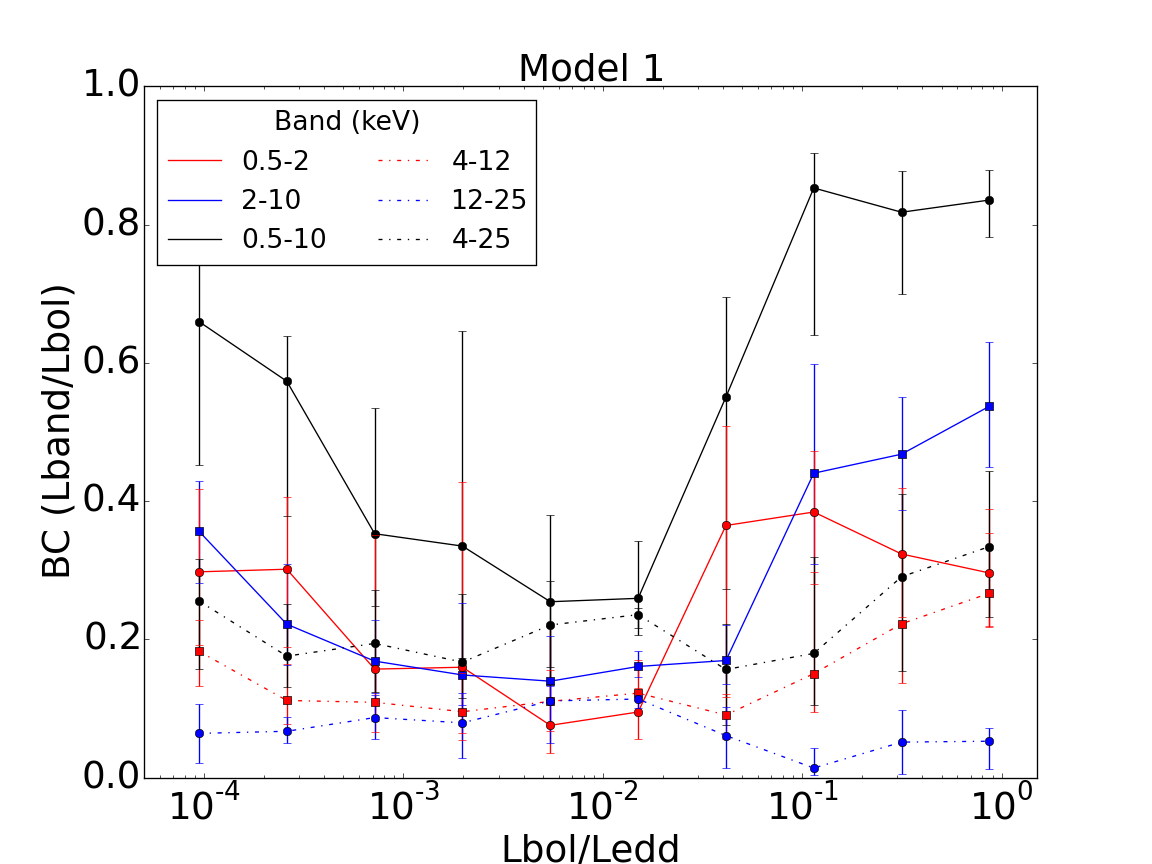}
		\caption{Average BCs for our sample of BHXBs described in Table \ref{tab.bhs} as a function of the Eddington ratio ($\mathrm{L_{bol}/L_{edd}}$). Here we present the BCs for spectral model 1 (see Section \ref{BHs}) and all energy bands described in Table \ref{tab.bands}.	Namely the dash-dotted lines represent the BCs for 4-12 keV, 12-25 keV, and 4-25 keV in red, blue, and black respectively. The solid lines represent the BCs for 0.5-2 keV, 2-10 keV, and 0.5-10 keV in red, blue, and black respectively.}
		\label{fig.bc_bh_model1}
\end{figure}

\begin{figure}
	\centering
		\includegraphics[width=1.0\columnwidth]{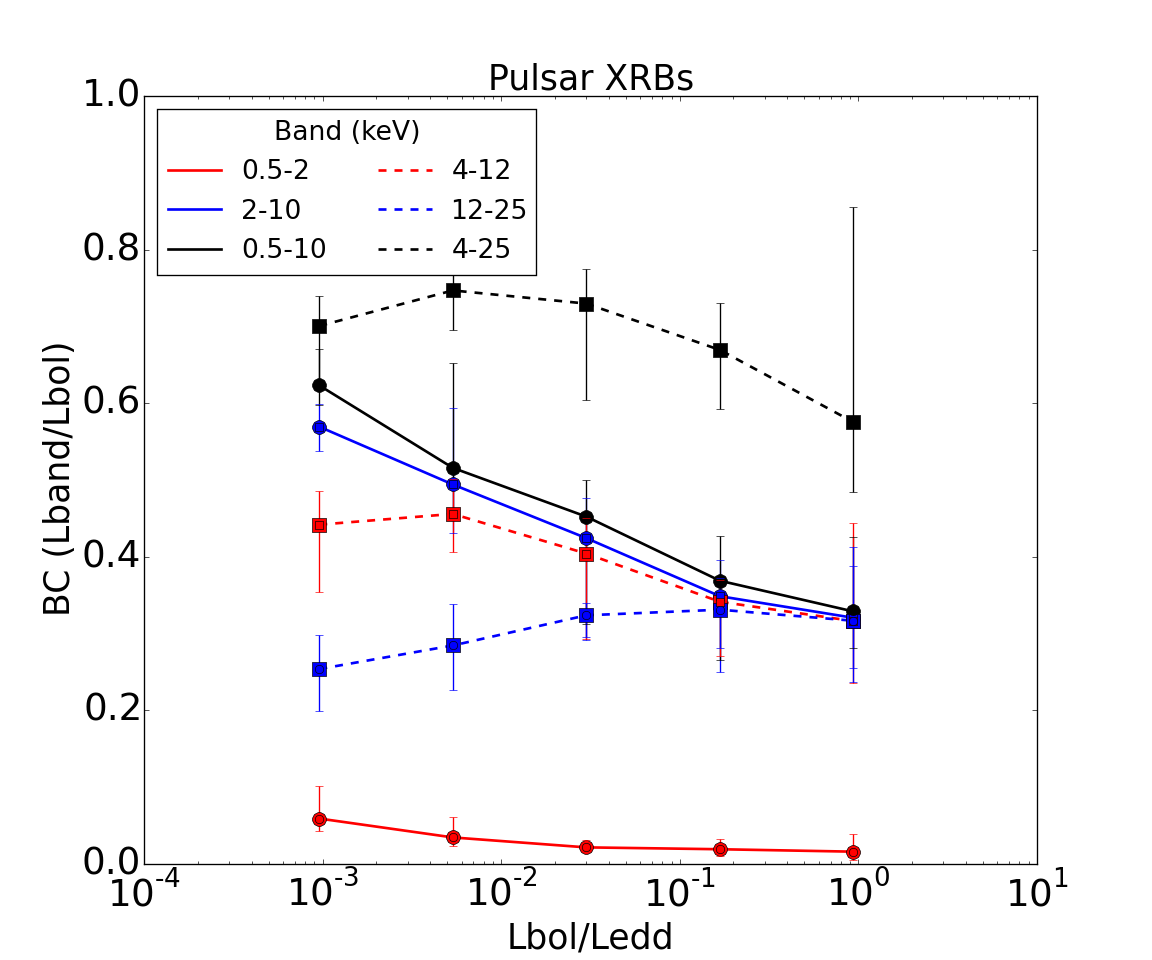}
		\caption{Average BCs for our sample of pulsar XRBs described in Table \ref{tab.pulsars} as a function of the Eddington ratio ($\mathrm{L_{bol}/L_{edd}}$), and assuming intrinsic \nh{}. The colour coding is the same as described in Fig.\ref{fig.bc_bh_model1}}
		\label{fig.pulsars_bc_intrisic}
\end{figure}

In Fig.\ref{fig.bc_bh_model1}, we show for model 1 and unabsorbed spectra of BHXBs how the average BC changes at different accretion rates.
BCs of bin 1 ($L_{bol}/L_{edd}\sim 10^{-4}$), should be interpreted with caution, since they have significant systematics related to the background subtraction.
The different colours correspond to BCs for different energy bands. For band 0.5-10 keV (solid black line) we notice that BCs are higher at lower and higher accretion rates. This is the result of a softer spectrum and a stronger accretion disk for higher accretion rates and a decrease in the hard X-ray flux at lower accretion rates. 
Therefore most of the 0.5-10 keV band emission dominates the spectrum at high accretion rates above 0.1 $L_{edd}$. Similar behaviour show the sub-bands 0.5-2 keV (solid red line) and 2-10 keV (solid blue line). However it is clear that the 2-10 keV contributes more to the higher BCs at Eddington ratios $>$0.1. 

The rest of the bands which correspond to higher energy parts of the spectrum, 4-25 keV (dotted black line), 4-12 keV (red dotted line), and 12-25 keV (blue dotted line) show similar behaviour. First, in all cases the BCs are quite low ($<0.3$ BC) with the lowest ($<0.1$ BC) for the harder band of the three, revealing that the contribution of the emission above 4 keV to the bolometric emission is rather low. The variation of the BC as a function of Eddington ratio is also not significant ($<$0.1 BC).
For band 4-25 keV, there is a small increase for low ($\sim 10^{-4}$), high ($\sim 1$) and intermediate ($\sim 10^{-2}$) Eddington ratios. These small increases could be explained by the different contributions in the 2 sub-bands. Namely, band 4-12 keV contributes to this increase for low ($\sim 10^{-4}$) and high ($\sim 1$) Eddington ratios, while band 12-25 keV contributes in the intermediate ($\sim 10^{-2}$) Eddington ratios.

Moreover, we observe (Fig. \ref{fig.bh_bc}) that the BCs for the different spectral models of BHXBs (described in section \ref{BHs}), show the same behaviour for all bands. Therefore, using a different model does not affect the BC calculation.
In Tables \ref{tab.bcbhs} to \ref{tab.bcbhs4}, we present the values of BCs for all available models of BHs XRBs at different Eddington ratio bins.

\subsubsection{Pulsars XRBs}

In Fig. \ref{fig.pulsars_bc_intrisic}, we show how the BC change as a function of the Eddington ratio for pulsar XRBs and intrinsic absorption (see section \ref{nh}).
Very interesting is the fact that the behaviour of BC as a function of Eddington ratio for pulsars, looks very similar with those of BHs XRBs, however for the higher accretion rates this is only true when we include the confirmed pulsar ULXs (see following Section).
Moreover for pulsars XRBs, we overall notice that at harder bands (4-25 keV, 4-12 keV, 12-25 keV), the BCs have higher values than the ones observed in BHs. This is expected since pulsar XRBs generally show harder spectra comparison to those of BHXBs \citep[e.g.][]{vrtilek13}.
We report the BC values for pulsar XRBs and intrinsic \nh{} in Table \ref{tab.bcpulsars_in}.

\subsubsection{ULXs}

\begin{figure*}
	
	\begin{minipage}{170mm}
	\centering
	  \includegraphics[width=0.325\columnwidth]{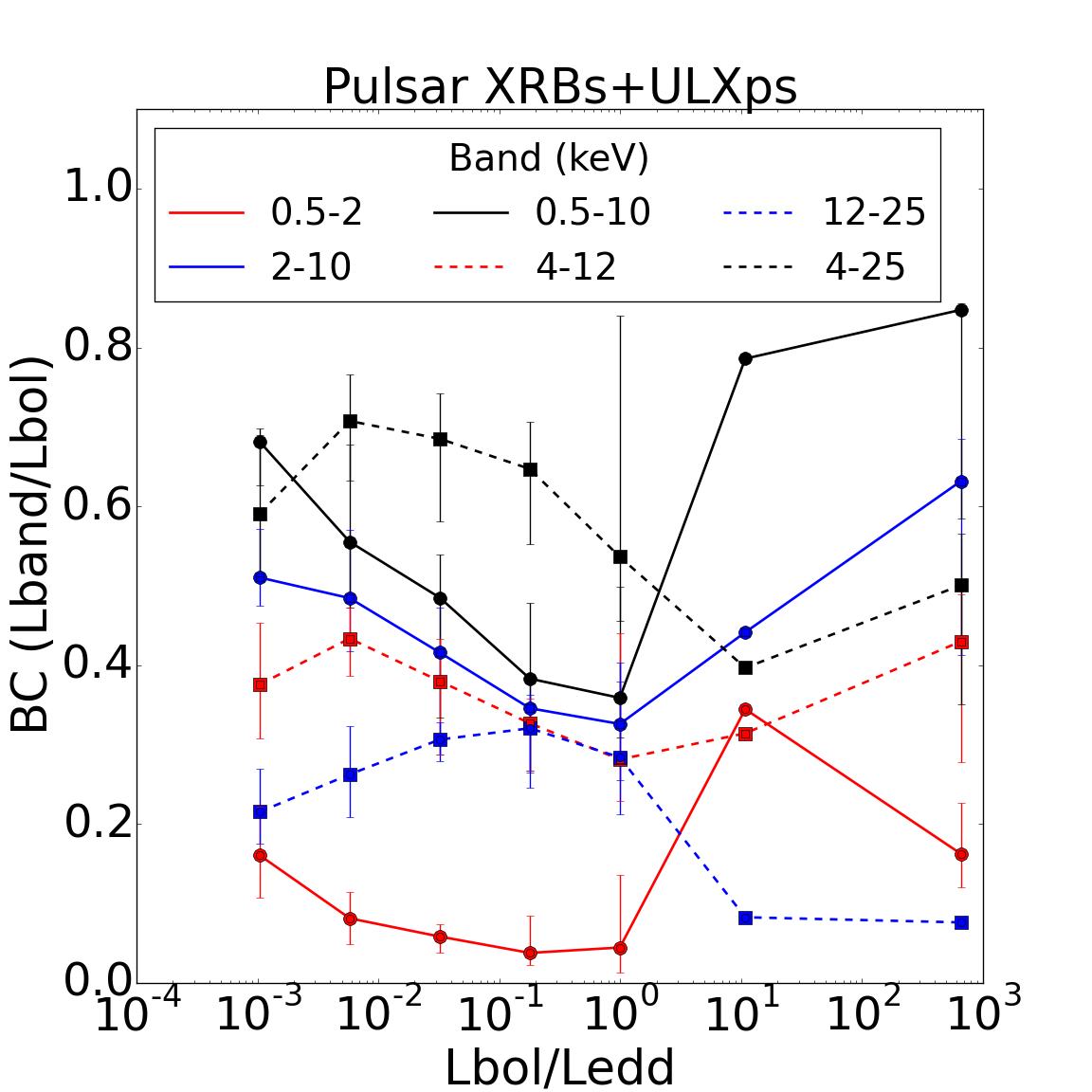}
	  \includegraphics[width=0.325\columnwidth]{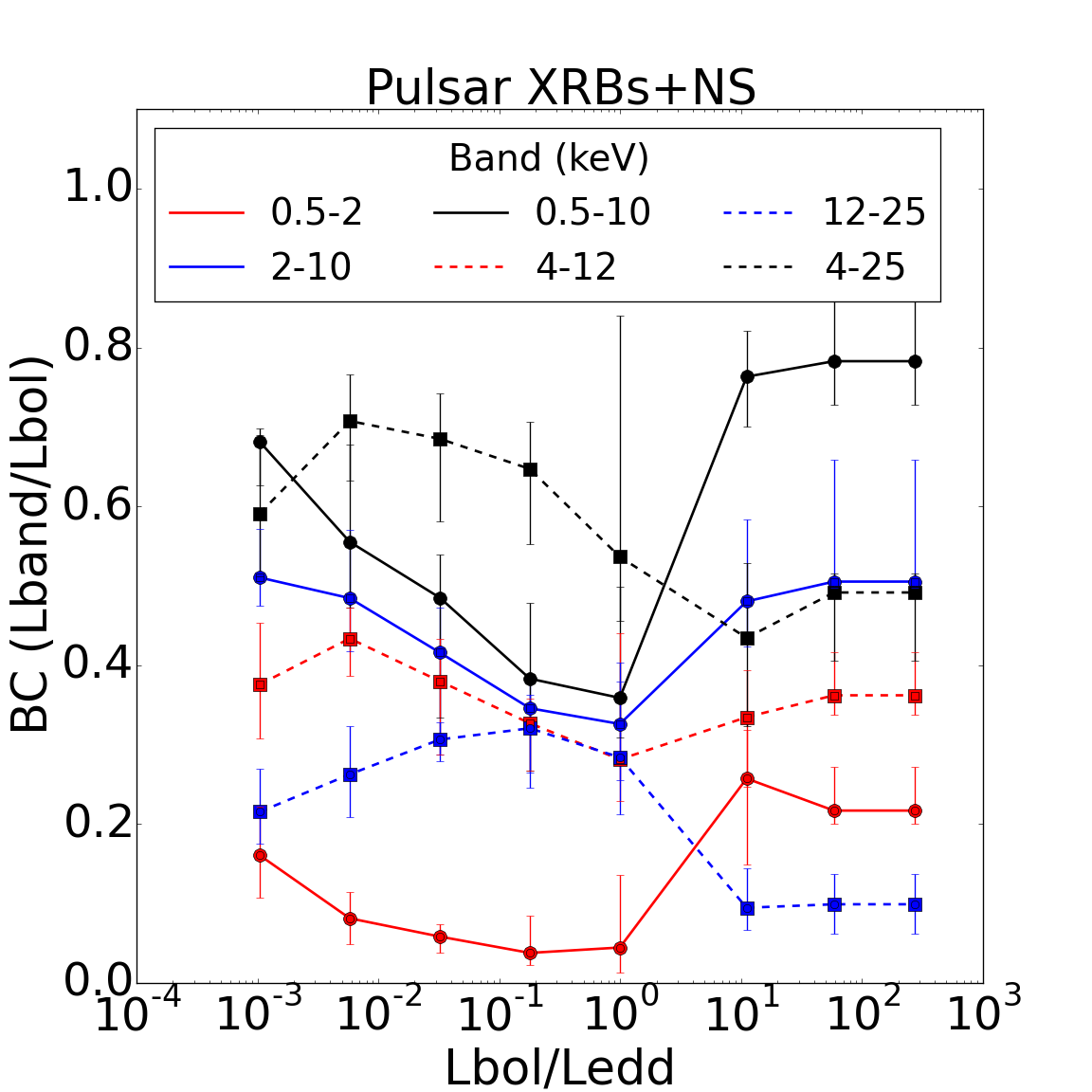}
	  	\includegraphics[width=0.325\columnwidth]{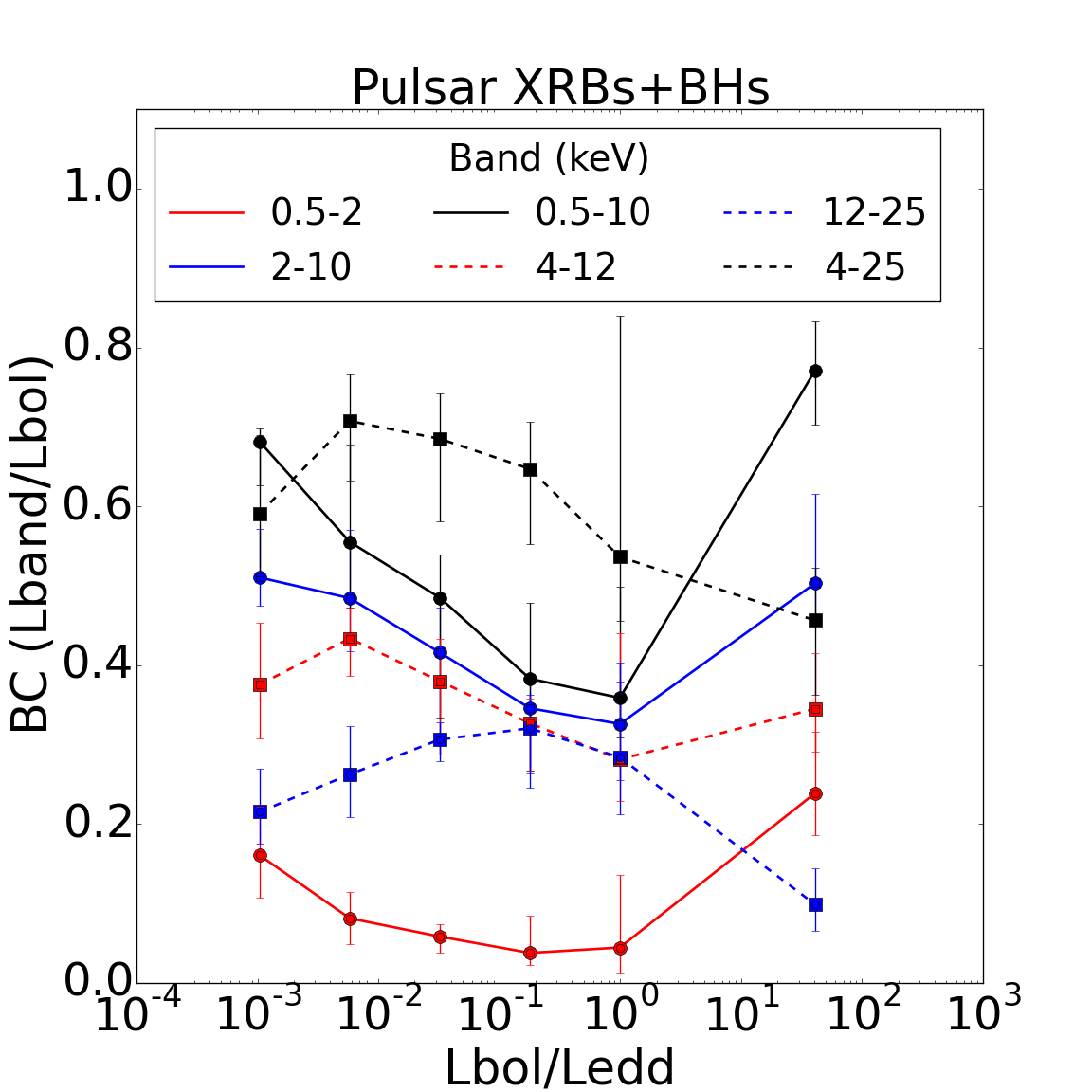} 
		\caption{
	Average BCs for our sample of pulsar XRBs and ULXs as described in Tables \ref{tab.pulsars} and \ref{tab.ulxs}, and as a function of the Eddington ratio ($\mathrm{L_{bol}/L_{edd}}$) and assuming fiducial \nh{}. From left to right plots, we account for the contribution of confirmed pulsar ULXs, then for ULXs assuming that they are all NS, and finally for ULXs assuming that they all BHs (excluding the confirmed pulsar ULXs). The colour coding is the same as described in Fig.\ref{fig.bc_bh_model1}. 
		}
		\label{fig.pulsars_ulxs}
	\end{minipage}
\end{figure*}

In Fig. \ref{fig.pulsars_ulxs}, we present together with the data for the pulsar XRBs also how the BC change at higher Eddington ratios ($>$1), if we add spectral models for all ULXs of Table \ref{tab.ulxs}. We consider three cases, 1) where all confirmed pulsar ULXs are added (left panel); 2) assuming that all ULXs are NS with masses of 1.4\,M$_{\odot}$ (middle panel) and 3) assuming that all non-confirmed pulsar ULXs are BHs with a mass of 9.1\,M$_{\odot}$ (right panel). We see that no matter the assumption on the nature of the ULXs, the behaviour is equivalent for the same Eddington ratios. Under the assumption however that all ULXs are NS, the correlation expands to higher Eddington ratios ($\sim10^{3}$).
For cases 1) and 2) where the ULXs are treated as NS we decided to plot separately the source NGC 300 ULX1 (6th bin; $\mathrm{L_{bol}/L_{edd}}\sim$ 10), due to its more complicated spectrum compare to other ULXps (see \citet{carpano18}. For this particular reason this bin has no errors in our plots, and also is the reason why the BC shows an abrupt increase for the band 0.5-2 keV.
We report the fiducial \nh{} BC values for pulsar, ULXps, ULXs assumed as NS, and ULXs assumed as BHs in Tables \ref{tab.bcpulsars}, \ref{tab.bculxps}, \ref{tab.bculxns}, and \ref{tab.bculxbh} respectively.

\subsection{BHXBs disk optical to X-ray luminosity as a function of accretion rate. }\label{optical_discussion}

In XRB studies physical quantities regarding the system are usually inferred by measuring the optical emission of the donor star in quiescence. 
In transient systems such as these, during outburst the donor-star emission is generally negligible compared to that of other components (disk, jet etc).
 
In BH HMXBs the donor star is so massive ($\gtrsim 10 M_{\odot}$), that dominates the optical X-ray output of the system \citep{vandenheuvel72,treves80}, while in BH LMXBs ($M_{donor} \lesssim 3 M_{\odot}$), optical emission is mainly originating from the outer accretion disk as a result of reprocessing of X-rays produced in the inner disk \citep[][]{cunningham76, vrtilek90}. Other processes might be contributing such as thermal emission from the companion star while the source is in quiescence \citep[e.g.][]{greene01, mikolajewska05}, intrinsic thermal emission from a viscously heated outer disk \citep[][]{shakura73,frank02}, as well as possibly jet contribution in the hard state although it is expected to be higher in the NIR (\citet{russell06}, for a review see \citet{fender06}). 

\begin{figure}
	\centering
	\includegraphics[width=1.0\columnwidth]{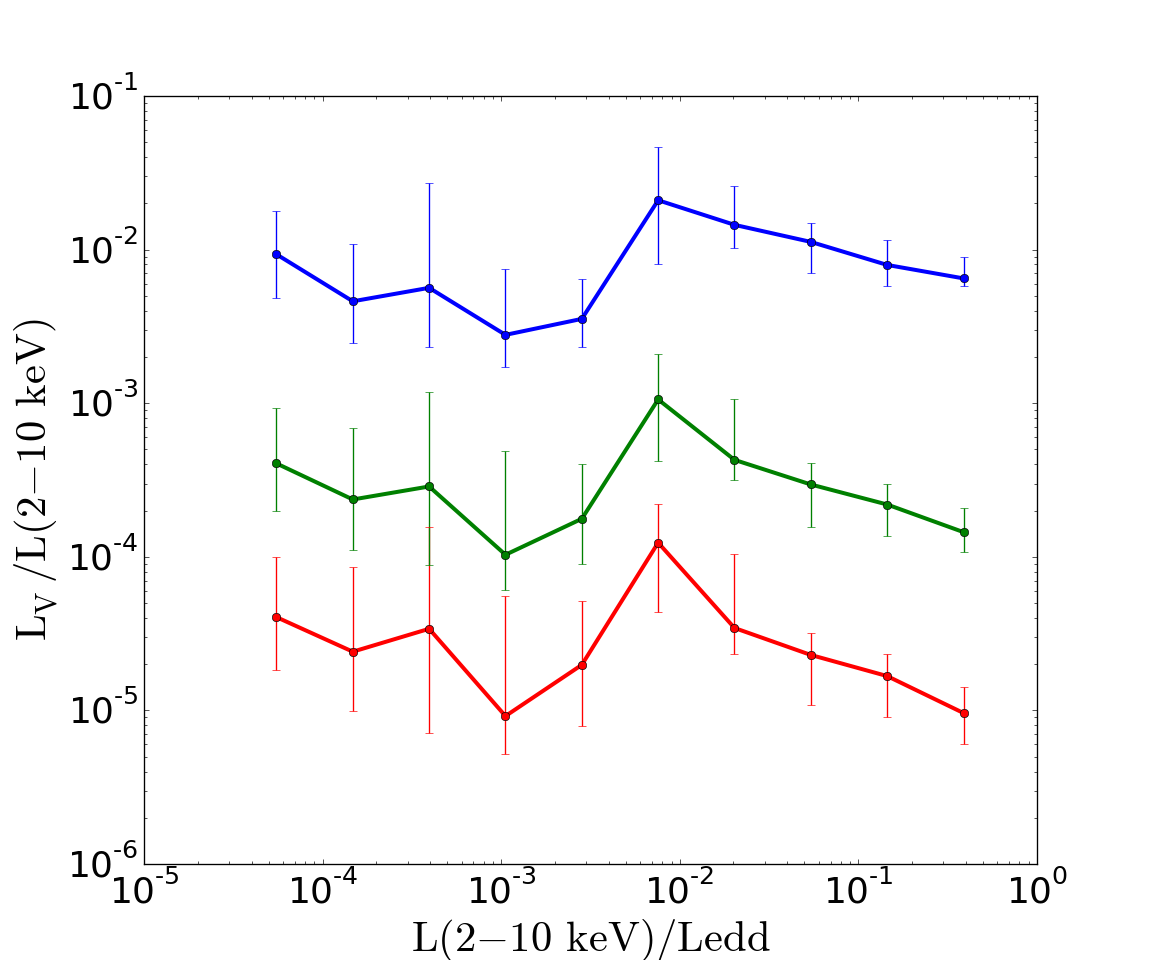}
	  		\caption{Disk optical V-band to X-ray 2-10 keV band luminosity ratio of the BHXRBs as a function of the Eddington ratio L(2-10 keV)/$L_{edd}$. Different colours correspond to different dependencies of the local disk temperature $T(r)$ to the radius $r$, i.e different disk geometries. Namely the red line corresponds to $T(r) \sim r^{-0.75}$ (XSPEC model: \texttt{diskbb}), the green and blue lines to $T(r) \sim r^{-0.65}$ and $T(r) \sim r^{-0.55}$ respectively (XSPEC model: \texttt{diskpbb}). }
		\label{fig.lxlopt}
\end{figure}

In this section we focus on the optical emission originating from X-ray reprocessing in the outer accretion disk and we are measuring its expected contribution at different accretion rates. In more detail, in Fig. \ref{fig.lxlopt} (red line) we determine the optical (V-band; $L_V$) to X-ray (2-10 keV; $L_X$) ratio originating from the disk/corona and we show how it changes as a function of Eddington ratio. Here we use the extrapolated fluxes from the models as described in Section \ref{optical}. We see that accounting for the 1$\sigma$ errors the $L_V/L_X$ value shows no significant variation ($L_V/L_X\sim 4\times 10^{-5}$) as a function of accretion rate except of a small increase at $L_X$/$L_{edd}$ $\sim 10^{-2}$, which reveals that independent of the state a binary is in (soft, hard, etc), and under the assumption that the contribution of the jet is negligible in the soft and not significant during the hard state and the optical band, the amount of the optical to X-ray emission remains constant as a function of Eddington ratio.

This result is not unexpected, since correlations between X-ray and optical emission have been identified. 
For example, \citet{russell06} using quasi simultaneous optical and X-ray observations from XRBs observed that the hard state near-infrared/optical luminosity of BH LMXBs (originating from the disk and the jet) increases with increasing 2-10 keV X-ray luminosity (see top panel of their Fig.1). Regarding the soft state, the optical luminosity (see bottom panel of their Fig.1) seems to increase as the X-ray luminosity is increasing more or less at the same rate as was observed in the hard state.  \citet{russell06} conclude also that the contribution of the jet is less important for the optical compared to the NIR band. This could give an explanation of why the optical luminosity increases with increasing X-ray luminosity almost at the same rate, for both the soft (where the jet is quenched) and the hard state. This result is is in agreement with a more or less constant X-ray to optical ratio luminosity we have calculated along different accretion rates (Fig. \ref{fig.lxlopt}).

\subsubsection{Testing different disk geometries}

The BH model spectra reported in this work, were fitted  using an XPSEC \texttt{diskbb} model to account for the X-ray emission originating from the accretion disk \citep{steiner16}. The diskbb model represents the spectrum from an accretion disk consisting of multiple black-body components where the local disk temperature is described as function of disk radius $T(r)\sim r^{-p}$ where $p=3/4$. This represents the standard picture of an optically thick geometrically thin accretion disk \citep[][]{shakura73}.

However studies have shown that radial advection could be important and then $p$ can have smaller values \citep[e.g.][]{mineshige94,watarai00,kubota04}. 
Another mechanism that could contribute to this effect is X-ray irradiation of the outer accretion disk by X-rays produced in the inner accretion flow \citep[e.g.][]{vanparadijs94,vanparadijs96}, which is believed to have a large impact on the outer disk during outburst.
Accretion disks with $p<0.75$ are described by the so-called slim accretion disk models which were proposed by \citet{abramowicz88} and represent optically thick, geometrically not very thin accretion disks. 
A different disk geometry could have significant implications on the reprocessing of X-rays in the disk and therefore on the optical emission originating from the disk.

We tested this effect by replacing in our spectral models for the BHXBs the \texttt{diskbb} component with a \texttt{diskpbb} component using two different values of $p$, namely $p=0.65$ and $p=0.55$. The substitution was made under the assumption that the 2-10 keV flux remains unchanged between the \texttt{diskbb} and \texttt{diskpbb} models, and then the V-band luminosity was calculated following the method in Section \ref{optical}.
We show in Fig. \ref{fig.lxlopt} how the $L_V/L_X$ changes as a function of accretion rate for different disk geometries. We observe that the optical flux increases significantly, one and two orders of magnitude for $p=0.65$ and $p=0.55$, respectively, therefore the thicker the disk geometry is, the higher is expected to be its optical emission.

\subsubsection{Comparison with observations}\label{observations}

The optical to X-ray luminosity ratio of the disk as function of different Eddington ratios and disk geometries as presented in Fig. \ref{fig.lxlopt} provides a diagnostic tool, when compared with observational data, not only to determine the disk's optical contribution in a BH XRB system at different accretion rates/spectral states, but also to indicate the geometry of the disk. For that reason we used a number of simultaneous and quasi-simultaneous optical and X-ray data of BHXBs available in the literature. 

We used the same distance and mass measurements reported in Table \ref{tab.bhs} for the sources that also exist in our sample, while for the rest, we used the measurements reported in the corresponding papers or in the \textit{BlackCAT} catalogue \citep{blackcat}.
In Fig. \ref{fig.lxlopt_observations} we present the same data as in Fig. \ref{fig.lxlopt} with the difference that on the y axis we show, the optical luminosity $L_V$, for easier comparison with the observational data which are also over-plotted.

In more detail, regarding simultaneous observations, we have used \swift{} \textit{XRT} and ultraviolet/optical (\textit{UVOT}) data from the 2005 outburst of the BH LMXB GRO J1655-40 \citep{brocksopp06}. We used the V magnitude, and the X-ray flux at 2-10 keV computed from \textit{XRT} spectral models of six of their observations \citep[Table 4 in][]{brocksopp06}, which correspond to the high/soft or very high state. We corrected the optical V magnitudes for reddening using the value of $A_V=3.7\pm 0.3$ reported in \citet{jonker04}.
We then calculated the V-band luminosity using the AB magnitude zeropoint, the distance of the source reported in Table (\ref{tab.bhs}), and the flux as $\lambda \ F_{\lambda}$ assuming that the spectral range of the filter can be approximated by its central wavelength.
In Fig. \ref{fig.lxlopt_observations} we show the data for GRO J1655-40 with magenta circles, which are located at Eddington ratios close to 1.0, which is generally expected for systems in high/soft or very high states, and between the green line for \texttt{diskpbb} model (p=0.65) and the blue line for the \texttt{diskpbb} model (p=0.55). The optical observations are not expected to include contribution from a jet, since it is considered to be quenched at the high/soft and very high spectral states \citet{fender06}. We calculated (using the spectral type absolute magnitude reported in \citet{carroll06}, and solar properties from \citet{prsa16}) that the the donor star, having a spectral type of F6IV (see Table \ref{tab.bhs}), is expected to have a contribution up to 14\% to the systems optical emission when the system is at $\sim0.05-0.1\ \mathrm{L_{edd}}$ (assuming the rest of the optical emission is originating from the accretion disk since the jet is quenched in the soft state). For that reason, we also corrected for the contribution of the donor star. We show in Fig. \ref{fig.lxlopt_observations}, that the disk optical luminosity of GRO J1655-40, agrees more with a what is expected from a thicker disk ($p\sim 0.60$).

We then included quasi-simultaneous observations of a number of the BHXBs presented in \citet{russell06}(see references therein). In Fig. \ref{fig.lxlopt_observations} we included data for BH LMXBs in the soft state (blue colour), in the hard state (red colour), as well as BH HMXBs (cyan colour), keeping always the same symbol for the same source. We should note here that the hard-state data taken from \citet{russell06} have been corrected for the fractional contribution of the donor star, and uncertainties connected with the uncertain spectral type of the donor star are propagated in the errors. Also they corrected for the contribution of the donor star, during outburst in cases the donor is comparatively bright, and as consequence all data points from \citet{russell06}, represent the optical flux originating from all other than the donor emission processes in the system. 

In more detail, regarding the data adopted from \citet{russell06}, we show the BH HMXBs Cyg-X1 (cyan circle) and V4641 Sgr (cyan star). We show the BH LMXBs LMC X-3 (square), 3A 0620-003 (pentagon), 4U 1543-475 (down triangle), V404 Cyg (big circle), GRO J1655-50 (x symbol), and XTE J1118+480 (plus symbol). 

All BH HMXBs are located above the blue line ($p=0.55$), and this is expected as the donor star dominates the optical output of the system. However the source V4641, could be characterised as an intermediate XRB being of spectral type B9III, as the $L_V$ contribution from the donor star is calculated (using absolute magnitude reported in \citet{carroll06}, and solar properties from \citet{prsa16}) to be of the order of $L_{V}\sim5.0 \times10^{35}$\ergs{} ,
which lies between 1\% and 5\% of the overall V-band luminosity of the system accounting for different observations which lie between 0.1 and 1.0 $\mathrm{L_{edd}}$ of the system. Therefore we consider more representative of the BH HMXB class, Cyg X-1. Overall, BH HMXBs are treated here as reference and an indication of a limit above which a source could be characterised as a BH HMXB.

Regarding the BH LMXBs, we notice that all sources regardless of the state they are in, they fall a bit below or above the blue line ($p=0.55$), which corresponds to the geometrically thicker disk case. 
In the soft state (blue points), the jet is quenched and therefore the optical emission we measure is dominated by X-ray emission and/or reprocessing from the disk.
V404 Cyg shows much higher optical luminosity than other soft BH LMXBs. This could be consequence of a very bright outburst it went through that completely alters the geometry of the disk, or the fact that it has a sub-giant companion.
In the hard state (red points), where the contribution of the donor star has been subtracted, we see only the contribution of the disk and the jet. However the contribution of the jet in the optical is not expected to be significant compared to that of the disk. In fact, \citet{russell06} showed that the transition between an optical thick to an optical thin jet happens around the K band, with the jet NIR contribution being $\sim90\%$.
They also measure that the optical contribution of the jet could be between zero and 76\% but the optical spectral energy distributions of BHXBs show a thermal spectrum indicating that X-ray reprocessing from the disk dominates the optical output.
Therefore significant contribution from the jet should be expected mainly in the hard state and in the NIR emission, while in the optical the disk dominates.
In all cases, all hard state data follow or are a bit above the blue line ($p=0.55$), in agreement with thicker disk geometries.
Moreover we see that none of the observed sources is below the red line ($p=0.75$), which is expected since no source could physically have an optical luminosity lower than the one produced by the disk at a given Eddington ratio.

\subsubsection{Discussion}

\begin{figure}
	\centering
	\includegraphics[width=1.0\columnwidth]{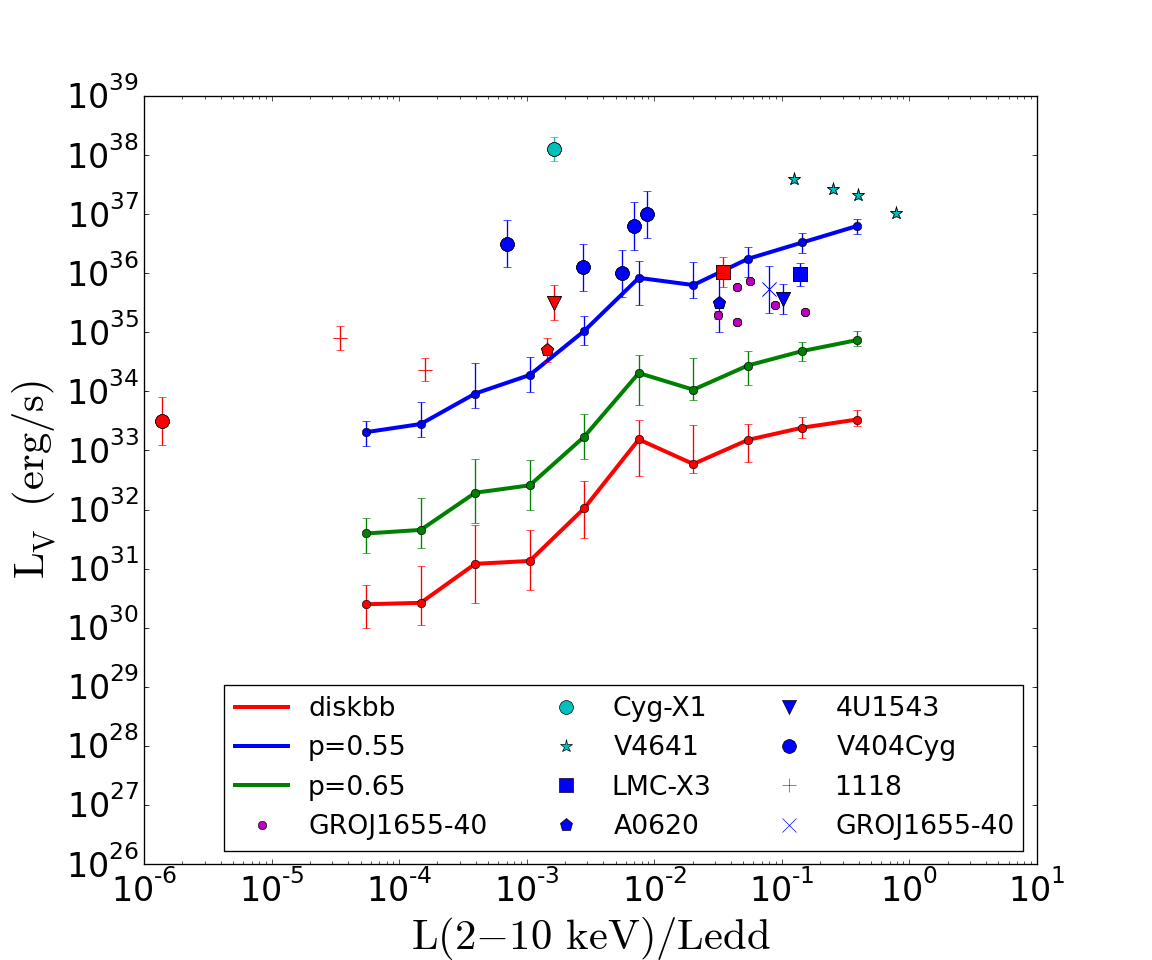}
	 		\caption{Optical V-band luminosity versus 2-10 keV band X-ray luminosity normalised by the Eddington luminosity. The solid lines represent the expected disk optical luminosity at different accretion rates and disk geometries of $p=0.75$, $p=0.65$, and $p=0.55$ with red, green, and blue respectively. The magenta circles correspond to simultaneous optical and X-ray observations of GRO J1655-40 \citep{brocksopp06}. 
	 		 Blue and red points correspond to quasi-simultaneous soft and hard state observations of XRBs while cyan to observations of BH HMXBs taken from \citet{russell06} and references therein. The same symbol is used for the same source.
	 		 }
		\label{fig.lxlopt_observations}
\end{figure}

Overall with the calculation of the $L_{V}/L_{X}$ as a function of Eddington ratio and the comparison with the observational data,  we aim to provide a limit of the disk optical contribution to the BH XRB system. This could be useful for a better determination of donor star type, or for estimating the disk geometry.
In fact we see that all observational data shown in Fig. \ref{fig.lxlopt_observations}, (under the assumption that the jet contribution in the hard state is not significant) do not agree with the standard scenario of a geometrically thin disk but rather with that of a geometrically thicker disk with $p\lesssim 0.60$.

However, there are certain assumptions included in our calculations that could introduce some level of uncertainty. For example, the spectral models we are using, assume an intrinsically steady state where the X-ray emission originating from the inner disk is coupled with the optical emission originating from the outer disk. However, two regimes of this behaviour are considered to exist, and our assumption is true in the regime where X-ray disk irradiation is dominant \citep[e.g.][]{vanparadijs94,vanparadijs96}. In this case, the emission at longer wavelengths corresponding to the outer portion of the disk is coupled to the emission produced in the inner disk. The second regime is dominated by viscous dissipation and therefore the outer disk optical emission and the inner disk X-ray emission are correlated with a time delay which depends on the viscous timescale, and in XRBs can be of the order of weeks \citep[e.g.][]{steiner14a}. In that latter case, our assumption would not hold, and any variability or any time delay effectively will add some noise on the relations we provide. However, since we do not use a time dependent model this point of uncertainty is not addressed to this work and could be let to some other kind of analysis that investigates the time dependent behaviour of disk models.

Another source of uncertainty introduced in our calculations is the contribution and effect of the corona into our results. On the one hand, the contribution of the corona is expected to be significant in the X-ray regime, and the subtraction of that contribution would result in somehow larger optical emission and therefore steeper relations. On the other hand, the corona could be in part responsible for the irradiation of the disk, resulting in higher optical emission and shallower relation. Disentangling the corona contribution is not possible in our calculations since it could potentially have an effect on both the X-ray and optical regime.
However, even though the coronal contribution, might be important in particular accretion regimes (i.e. very high state, low state), in the soft state  where a source spends most of its time during outburst \citep[e.g.][]{tetarenko16} it is not the dominant mechanism.
Therefore we do not expect that the coronal contribution would have significant implications into our results.

Interestingly, the fact that all of the observational data agree with a thicker disk scenario ($p\lesssim 0.60$), seems to be also in agreement with the lack of high inclination systems in transient BHXBs with low-mass companions \citep{narayan05,corralsantana13}. Although high-mass BHXBs can exhibit eclipses even at relatively low inclinations \citep[e.g. M33 X-7;][]{orosz07}, for low-mass BHXBs a lack of eclipses has been observed which can be attributed solely on inclinations effects.
For example, \citet{narayan05} after studying 11 sources with inclination estimates, found that no source has $i>75^{\circ}$. They showed that the probability that this would happen by chance is less than 4\%, and the lack of such systems is possibly the result of a flared accretion disk (corrugated or warped) by about $\pm15\ \mathrm{degrees}$.
Moreover they found that BHXBs with higher inclinations $i=70-75\ \mathrm{degrees}$ show more noisy light-curves, which could be explained if they are close to be occulted by the disk.
Therefore a flared thicker disk could be an explanation, for not seeing perfect eclipses in low-mass black-hole XRBs and as a consequence missing a large number of high-inclination systems.

\section{Summary}\label{summary}
In this paper we use a spectral model library of 10 BHXBs (9237 models), 9 pulsar XRBs (577 models) and 13 ULXs (32 models).
Using this library we calculated:
\begin{enumerate}
\item Bolometric corrections in six energy bands (0.5-2 keV, 2-10 keV, 0.5-10 keV, 4-12 keV, 12-25 keV, and 4-25 keV) as a function of different Eddington ratios. We cover Eddington ratios from $\sim 10^{-4}\ L_{edd}$ for the BH and pulsar XRBs, and reaching up to $\sim 10^{3}\ L_{edd}$ when including the ULXs to study the higher-end of Eddington ratios.
We notice that the same pattern is observed, shifted by $\sim 2$ orders in the Eddington ratio, between the BCs calculated for BH and pulsar XRB, only when we include the pulsar ULXs in the latter case.
Moreover fitting with different spectral models for BHs does not affect the BC calculations.

Average BCs as a function of the accretion rate, are useful tools for calculating the bolometric luminosity of any XRB using its luminosity in the narrow band provided by the existing X-ray observatories, and in calculations for XRB population synthesis models as observable constraints.

\item For BHXRBs, optical to X-ray luminosity ratio as a function of Eddington ratio originating from the accretion disk and corona, and for different disk geometries ($p=0.55$,$p=0.65$, and $p=0.75$). We observe that the optical to X-ray contribution for the same disk geometry does not show any significant variation, while as we move to thicker disks (smaller $p$ values), the optical contribution of the disk to the binary system increases.
Including quasi-simultaneous optical and X-ray observational data, which are representative of the disk emission, we find that they agree with a thicker disk scenario instead of the standard thin disk. The existence of a disk thicker than the standard one ($p<0.75$) could also explain the fact that we do not observe any high inclination low-mass BH systems.

Overall, the knowledge of the average optical contribution from the disk as a function of the accretion rate, and for different geometries, is useful for a better determination of the spectral type of the donor star (after exclusion of the disk luminosity), as well as for the estimation of the disk geometry. 
\end{enumerate}

\section*{Acknowledgements}

K. A. and A. Z. acknowledge funding from the European Research Council under the European Union's Seventh Framework Programme (FP/2007-2013)/ERC Grant Agreement n. 617001 and from the European Union's Horizon 2020 research and innovation program me under the Marie Sklodowska-Curie RISE action, grant agreement No 691164 (ASTROSTAT). K. A. acknowledges funding from the European
Research Council (ERC) under the European Union’s Horizon
2020 research and innovation programme (grant agreement No
865637). We also made use of NASA’s Astrophysics Data System Bibliographic Services.

\section*{Data availability}
Spectra models of BH and pulsar XRBs used in this work, resulted from spectra extracted by public data of the \textit{RXTE} observatory. Spectral models used for ULXs are reported in the corresponding references.




\appendix

\section{Tables and plots}

 \begin{figure*}
 	
 	\begin{minipage}{160mm} 
 	\centering
 	 \includegraphics[width=0.6\textwidth]{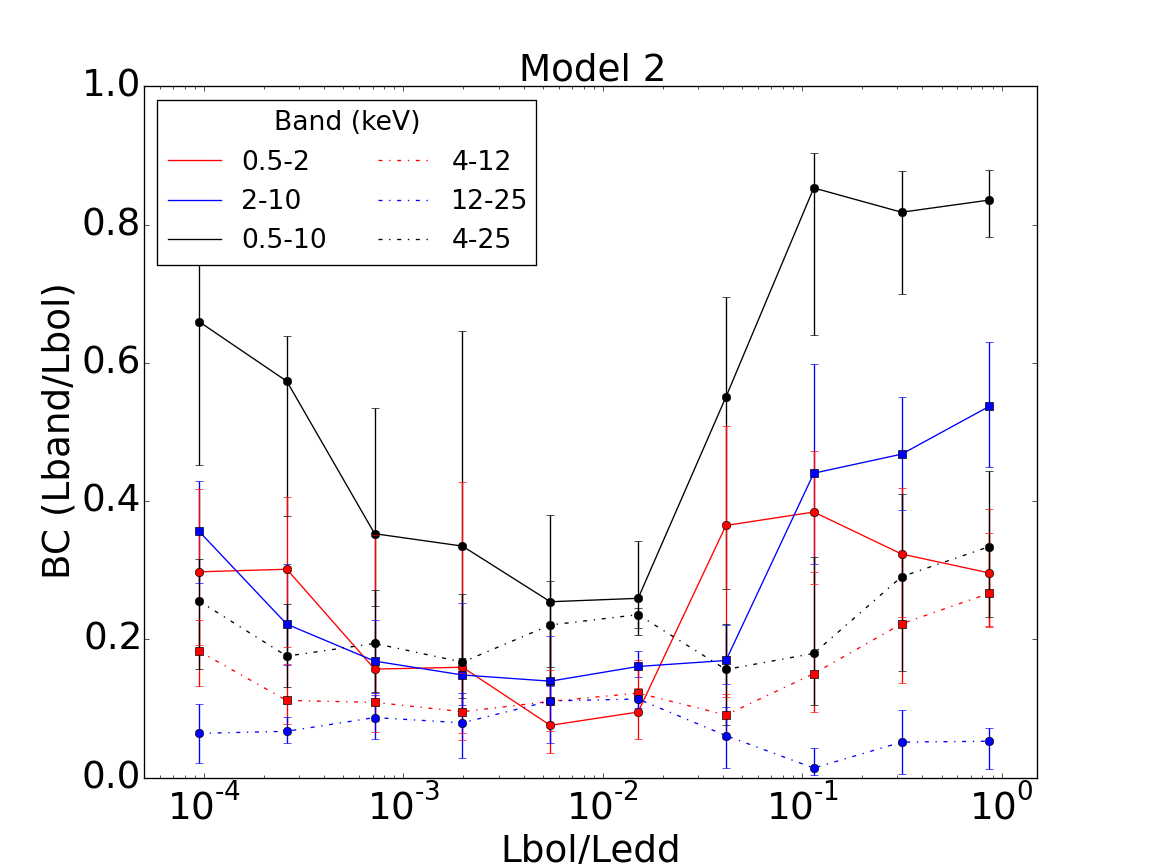}
 	 \includegraphics[width=0.6\textwidth]{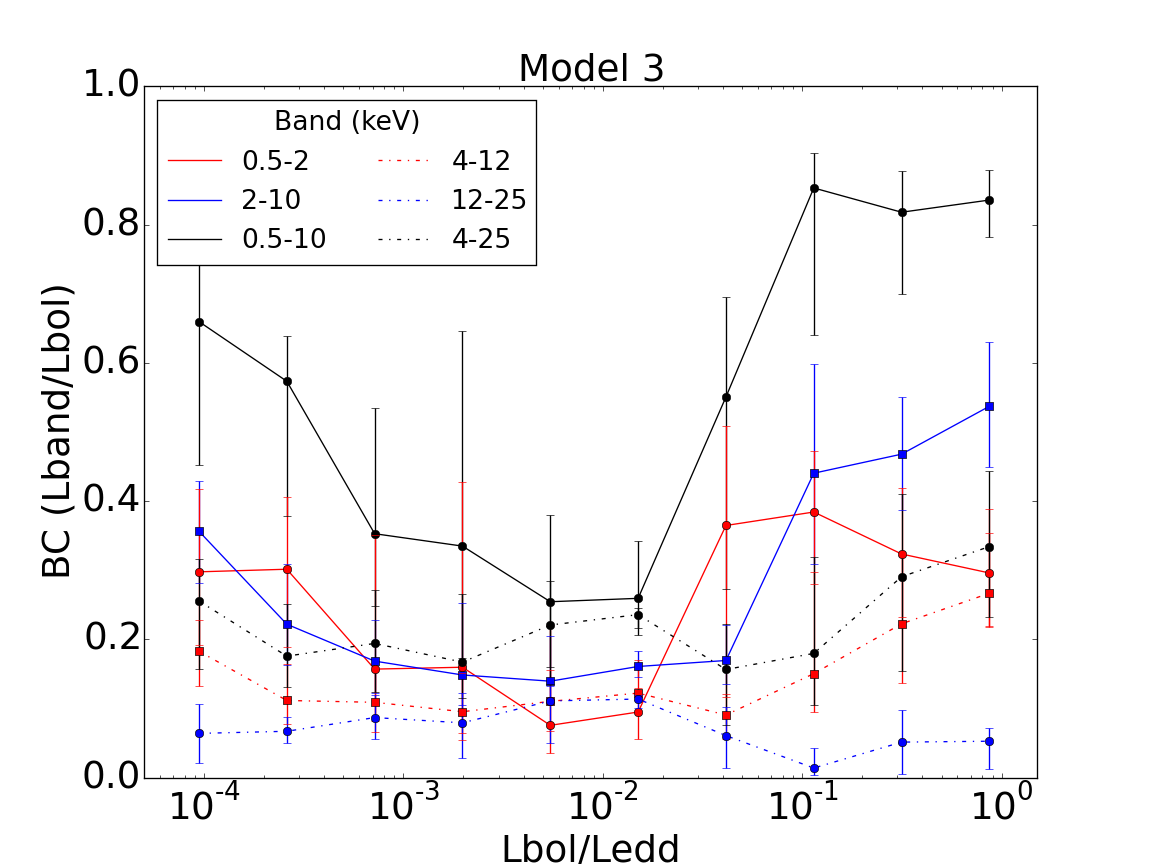}
 	 	\includegraphics[width=0.6\textwidth]{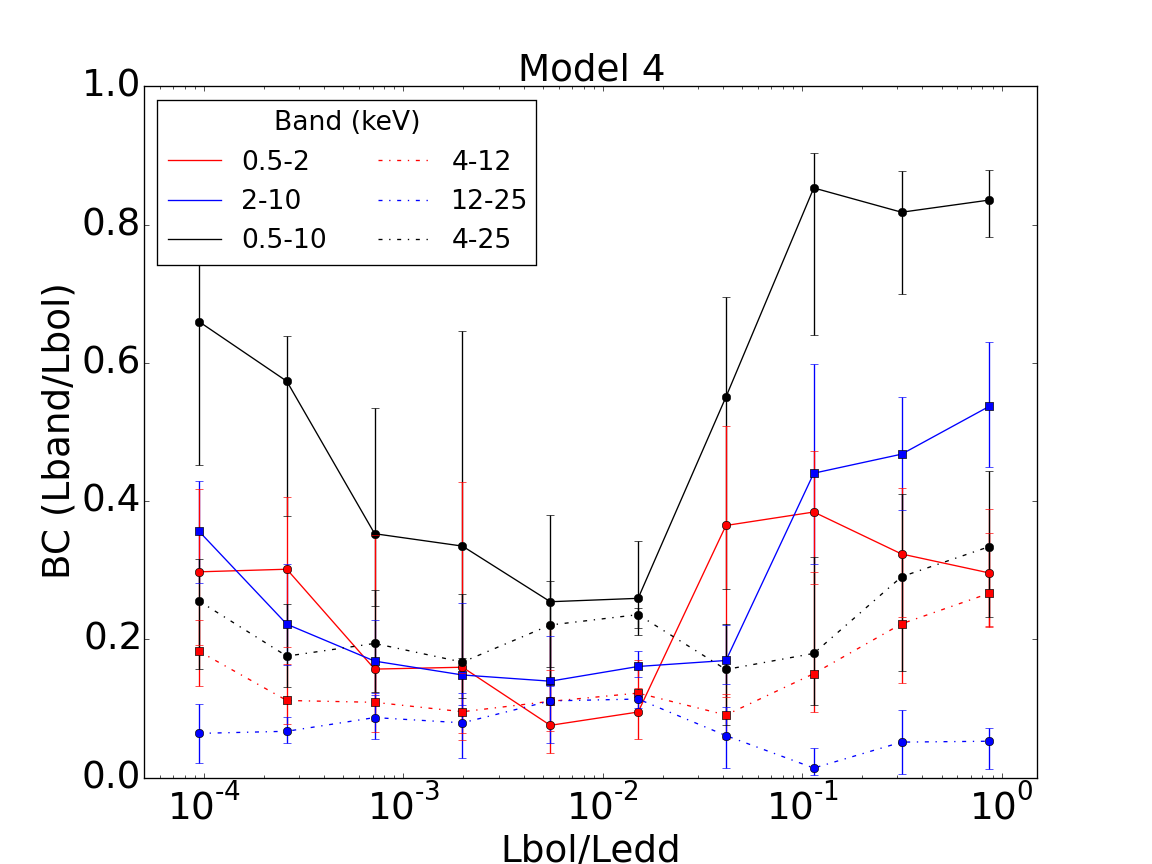} 
 		\caption{Average BCs for our sample of BHXBs described in Table \ref{tab.bhs} as a function of the Eddington ratio ($\mathrm{L_{bol}/L_{edd}}$). Here we present the BCs for spectral models 2, 3, and 4 (see Section \ref{BHs}), and all energy bands described in Table \ref{tab.bands} : Model 2 in the top, model 3 in the middle, and model 4 in the bottom panel. Color coding is the same as in Fig. \ref{fig.bc_bh_model1}. }
 		\label{fig.bh_bc}
 	\end{minipage}
 \end{figure*}

 \begin{table*}
		\begin{minipage}{140mm}
		\centering
		\caption{Average BCs for model 1 of the BHXBs in 10 Eddington ratio bins.}
		\begin{tabular}{@{}ccccccccc@{}}
			\hline 
			N& $\log_{min} \frac{L_{bol}}{L_{edd}}$ &$\log_{max} \frac{L_{bol}}{L_{edd}}$& S1 & H1& B1 & S2& H2 & B2 \\
			\hline
1 & -4.3 &-3.86  & 0.3$_{-0.11}^{+0.12}$ & 0.36$\pm0.07$ & 0.66$_{-0.21}^{+0.14}$ & 0.18$\pm0.05$ & 0.06$\pm0.04$ & 0.26$_{-0.1}^{+0.06}$ \\[5pt]
2 & -3.86& -3.42  & 0.3$_{-0.11}^{+0.1}$ & 0.22$_{-0.06}^{+0.09}$ & 0.57$_{-0.19}^{+0.07}$ & 0.11$_{-0.03}^{+0.05}$ & 0.07$\pm0.02$ & 0.18$_{-0.04}^{+0.08}$ \\[5pt] 
3 & -3.42& -2.98 	 & 0.16$_{-0.08}^{+0.19}$ & 0.17$_{-0.05}^{+0.06}$ & 0.35$_{-0.11}^{+0.18}$ & 0.11$_{-0.04}^{+0.05}$ & 0.09$\pm0.03$ & 0.19$_{-0.07}^{+0.08}$ \\[5pt] 
4 & -2.98& -2.54 	 & 0.16$_{-0.11}^{+0.27}$ & 0.15$_{-0.04}^{+0.1}$ & 0.33$_{-0.16}^{+0.31}$ & 0.1$_{-0.03}^{+0.06}$ & 0.08$_{-0.05}^{+0.04}$ & 0.17$_{-0.05}^{+0.1}$ \\[5pt] 
5 & -2.54& -2.1 & 0.08$_{-0.04}^{+0.14}$ & 0.14$_{-0.03}^{+0.07}$ & 0.25$_{-0.09}^{+0.13}$ & 0.11$_{-0.04}^{+0.05}$ & 0.11$_{-0.06}^{+0.02}$ & 0.22$_{-0.09}^{+0.06}$ \\[5pt] 
6 & -2.1 &-1.66  & 0.09$_{-0.04}^{+0.08}$ & 0.16$\pm0.02$ & 0.26$_{-0.05}^{+0.08}$ & 0.12$\pm0.01$ & 0.11$\pm0.01$ & 0.24$_{-0.02}^{+0.01}$ \\[5pt] 
7 & -1.66& -1.22 & 0.36$_{-0.25}^{+0.14}$ & 0.17$_{-0.03}^{+0.05}$ & 0.55$_{-0.28}^{+0.14}$ & 0.09$\pm0.03$ & 0.06$_{-0.05}^{+0.04}$ & 0.16$_{-0.08}^{+0.07}$ \\[5pt] 
8 & -1.22& -0.78 & 0.38$_{-0.1}^{+0.09}$ & 0.44$_{-0.13}^{+0.16}$ & 0.85$_{-0.21}^{+0.05}$ & 0.15$_{-0.06}^{+0.15}$ & 0.01$_{-0.01}^{+0.03}$ & 0.18$_{-0.08}^{+0.14}$ \\[5pt] 
9 & -0.78& -0.34 & 0.32$_{-0.09}^{+0.1}$ & 0.47$\pm0.08$ & 0.82$_{-0.12}^{+0.06}$ & 0.22$_{-0.09}^{+0.1}$ & 0.05$\pm0.05$ & 0.29$_{-0.14}^{+0.12}$ \\[5pt] 
10& -0.34& 0.1 & 0.3$_{-0.08}^{+0.06}$ & 0.54$\pm0.09$ & 0.84$_{-0.05}^{+0.04}$ & 0.27$_{-0.05}^{+0.12}$ & 0.05$_{-0.04}^{+
0.02}$ & 0.33$_{-0.1}^{+0.11}$ \\[5pt] \hline
\end{tabular} 	
		\label{tab.bcbhs}
		\smallskip
		{}
	\end{minipage}
	\end{table*}

 \begin{table*}
	\centering
		\begin{minipage}{140mm}
		\caption{Average BCs for model 2 of the BHXBs in 10 Eddington ratio bins.}
		\begin{tabular}{@{}ccccccc@{}}
			\hline 
			N& S1 & H1& B1 & S2& H2 & B2 \\
			\hline
1&0.33$_{-0.14}^{+0.09}$ & 0.34$_{-0.05}^{+0.07}$ & 0.69$_{-0.17}^{+0.08}$ & 0.19$_{-0.06}^{+0.05}$ & 0.08$\pm0.04$ & 0.26$_{-0.1}^{+0.13}$ \\[5pt] 
2&0.3$_{-0.13}^{+0.1}$ & 0.3$_{-0.09}^{+0.05}$ & 0.61$_{-0.14}^{+0.09}$ & 0.16$_{-0.05}^{+0.06}$ & 0.09$_{-0.03}^{+0.05}$ & 0.25$\pm0.09$ \\[5pt] 
3&0.15$_{-0.07}^{+0.13}$ & 0.2$_{-0.06}^{+0.12}$ & 0.39$_{-0.15}^{+0.17}$ & 0.13$_{-0.06}^{+0.1}$ & 0.11$_{-0.04}^{+0.07}$ & 0.24$_{-0.1}^{+0.17}$ \\[5pt]
4&0.16$_{-0.1}^{+0.23}$ & 0.17$_{-0.07}^{+0.14}$ & 0.37$_{-0.19}^{+0.3}$ & 0.11$_{-0.05}^{+0.09}$ & 0.1$_{-0.05}^{+0.07}$ & 0.22$_{-0.1}^{+0.14}$ \\[5pt] 
5&0.09$_{-0.05}^{+0.21}$ & 0.15$_{-0.04}^{+0.11}$ & 0.29$_{-0.12}^{+0.19}$ & 0.12$_{-0.06}^{+0.09}$ & 0.12$_{-0.07}^{+0.06}$ & 0.23$_{-0.12}^{+0.14}$ \\[5pt] 
6&0.1$_{-0.04}^{+0.07}$ & 0.16$\pm0.02$ & 0.26$_{-0.05}^{+0.09}$ & 0.12$\pm0.01$ & 0.12$\pm0.01$ & 0.24$_{-0.02}^{+0.01}$ \\[5pt] 
7&0.38$_{-0.26}^{+0.13}$ & 0.17$_{-0.04}^{+0.07}$ & 0.57$_{-0.29}^{+0.14}$ & 0.09$\pm0.03$ & 0.06$_{-0.04}^{+0.05}$ & 0.15$\pm0.08$ \\[5pt] 
8&0.4$_{-0.11}^{+0.07}$ & 0.44$_{-0.11}^{+0.16}$ & 0.86$_{-0.22}^{+0.04}$ & 0.15$_{-0.05}^{+0.14}$ & 0.01$_{-0.01}^{+0.03}$ & 0.18$_{-0.07}^{+0.14}$ \\[5pt] 
9&0.34$_{-0.1}^{+0.08}$ & 0.46$_{-0.08}^{+0.09}$ & 0.81$_{-0.12}^{+0.07}$ & 0.21$_{-0.07}^{+0.1}$ & 0.06$_{-0.05}^{+0.04}$ & 0.28$_{-0.12}^{+0.12}$ \\[5pt] 
10&0.35$_{-0.1}^{+0.1}$ & 0.48$_{-0.19}^{+0.09}$ & 0.8$_{-0.08}^{+0.07}$ & 0.23$_{-0.1}^{+0.1}$ & 0.05$_{-0.04}^{+0.02}$ & 0.28$_{-0.11}^{+0.13}$ \\[5pt] 
\hline
 \end{tabular} 	
		\label{tab.bcbhs2}
		\smallskip
		{}
	\end{minipage}
	\end{table*}

\begin{table*}
	\centering
		\begin{minipage}{140mm}
		\caption{Average BCs for model 3 of the BHXBs in 10 Eddington ratio bins.}
		\begin{tabular}{@{}ccccccc@{}}
			\hline 
		N&	 S1 & H1& B1 & S2& H2 & B2 \\
			\hline
 1&0.24$_{-0.12}^{+0.08}$ & 0.25$_{-0.06}^{+0.13}$ & 0.48$_{-0.11}^{+0.26}$ & 0.15$_{-0.07}^{+0.06}$ & 0.07$_{-0.03}^{+0.05}$ & 0.21$_{-0.07}^{+0.1}$ \\[5pt] 
2&0.28$_{-0.08}^{+0.13}$ & 0.22$_{-0.07}^{+0.09}$ & 0.54$_{-0.16}^{+0.12}$ & 0.11$_{-0.04}^{+0.05}$ & 0.06$_{-0.02}^{+0.03}$ & 0.17$_{-0.04}^{+0.09}$ \\[5pt] 
3&0.18$_{-0.09}^{+0.19}$ & 0.15$_{-0.04}^{+0.08}$ & 0.35$_{-0.12}^{+0.21}$ & 0.09$_{-0.03}^{+0.05}$ & 0.08$\pm0.03$ & 0.17$_{-0.06}^{+0.08}$ \\[5pt] 
4&0.17$_{-0.1}^{+0.24}$ & 0.14$_{-0.05}^{+0.1}$ & 0.33$_{-0.16}^{+0.31}$ & 0.08$_{-0.02}^{+0.07}$ & 0.07$\pm0.04$ & 0.15$_{-0.05}^{+0.1}$ \\[5pt] 
5&0.08$_{-0.04}^{+0.22}$ & 0.14$_{-0.03}^{+0.07}$ & 0.26$_{-0.1}^{+0.14}$ & 0.11$_{-0.03}^{+0.05}$ & 0.11$_{-0.05}^{+0.02}$ & 0.22$_{-0.08}^{+0.06}$ \\[5pt] 
6&0.08$_{-0.03}^{+0.07}$ & 0.16$_{-0.02}^{+0.03}$ & 0.25$_{-0.05}^{+0.08}$ & 0.13$\pm0.01$ & 0.12$_{-0.01}^{+0.0}$ & 0.24$\pm0.01$ \\[5pt] 
7&0.41$_{-0.3}^{+0.1}$ & 0.18$_{-0.04}^{+0.05}$ & 0.6$_{-0.33}^{+0.11}$ & 0.09$\pm0.03$ & 0.05$_{-0.04}^{+0.05}$ & 0.14$_{-0.07}^{+0.08}$ \\[5pt] 
8&0.38$_{-0.11}^{+0.09}$ & 0.43$_{-0.13}^{+0.17}$ & 0.85$_{-0.21}^{+0.05}$ & 0.15$_{-0.06}^{+0.15}$ & 0.01$_{-0.01}^{+0.03}$ & 0.18$_{-0.08}^{+0.14}$ \\[5pt] 
9&0.32$_{-0.08}^{+0.11}$ & 0.46$_{-0.07}^{+0.09}$ & 0.82$_{-0.12}^{+0.06}$ & 0.23$\pm0.09$ & 0.05$_{-0.05}^{+0.04}$ & 0.3$_{-0.15}^{+0.11}$ \\[5pt] 
10&0.29$_{-0.07}^{+0.08}$ & 0.53$_{-0.08}^{+0.1}$ & 0.84$_{-0.05}^{+0.04}$ & 0.28$_{-0.07}^{+0.12}$ & 0.05$_{-0.04}^{+0.02}$ & 0.34$_{-0.12}^{+0.11}$ \\[5pt]
 \hline
 \end{tabular} 	
		\label{tab.bcbhs3}
		\smallskip
		{}
	\end{minipage}
	\end{table*}

\begin{table*}
	\centering
		\begin{minipage}{140mm}
		\caption{Average BCs for model 4 of the BHXBs in 10 Eddington ratio bins.}
		\begin{tabular}{@{}ccccccc@{}}
			\hline 
		N&	 S1 & H1& B1 & S2& H2 & B2 \\
			 \hline
1 & 0.3$_{-0.1}^{+0.13}$ & 0.33$\pm0.08$ & 0.67$_{-0.21}^{+0.08}$ & 0.19$_{-0.09}^{+0.04}$ & 0.07$\pm0.04$ & 0.25$_{-0.13}^{+0.09}$ \\[5pt] 
2 &0.34$_{-0.18}^{+0.09}$ & 0.24$\pm0.06$ & 0.59$_{-0.19}^{+0.08}$ & 0.13$_{-0.04}^{+0.06}$ & 0.07$\pm0.03$ & 0.2$_{-0.05}^{+0.09}$ \\[5pt] 
3 &0.15$_{-0.07}^{+0.23}$ & 0.19$_{-0.04}^{+0.05}$ & 0.37$_{-0.1}^{+0.2}$ & 0.13$_{-0.05}^{+0.03}$ & 0.1$_{-0.04}^{+0.03}$ & 0.23$_{-0.08}^{+0.06}$ \\[5pt] 
4 &0.19$_{-0.13}^{+0.29}$ & 0.18$_{-0.05}^{+0.1}$ & 0.38$_{-0.16}^{+0.3}$ & 0.12$\pm0.05$ & 0.1$_{-0.08}^{+0.03}$ & 0.22$_{-0.12}^{+0.07}$ \\[5pt] 
5 &0.07$_{-0.04}^{+0.25}$ & 0.15$_{-0.04}^{+0.06}$ & 0.26$_{-0.11}^{+0.17}$ & 0.11$_{-0.04}^{+0.05}$ & 0.11$_{-0.06}^{+0.02}$ & 0.22$_{-0.08}^{+0.07}$ \\[5pt] 
6 &0.05$_{-0.01}^{+0.07}$ & 0.15$\pm0.03$ & 0.2$_{-0.02}^{+0.07}$ & 0.12$_{-0.02}^{+0.01}$ & 0.11$\pm0.01$ & 0.23$_{-0.04}^{+0.02}$ \\[5pt] 
7 &0.39$_{-0.27}^{+0.11}$ & 0.16$_{-0.04}^{+0.13}$ & 0.56$_{-0.3}^{+0.22}$ & 0.09$_{-0.03}^{+0.02}$ & 0.05$_{-0.04}^{+0.05}$ & 0.15$\pm0.06$ \\[5pt] 
8 &0.39$_{-0.11}^{+0.08}$ & 0.43$_{-0.11}^{+0.18}$ & 0.86$_{-0.22}^{+0.05}$ & 0.15$_{-0.05}^{+0.15}$ & 0.01$_{-0.01}^{+0.03}$ & 0.18$_{-0.08}^{+0.14}$ \\[5pt] 
9 &0.32$\pm0.09$ & 0.48$_{-0.08}^{+0.07}$ & 0.82$_{-0.12}^{+0.06}$ & 0.23$_{-0.08}^{+0.11}$ & 0.05$\pm0.05$ & 0.3$_{-0.14}^{+0.12}$ \\[5pt] 
10 &0.3$_{-0.08}^{+0.05}$ & 0.54$_{-0.09}^{+0.1}$ & 0.84$_{-0.06}^{+0.04}$ & 0.26$_{-0.04}^{+0.14}$ & 0.05$_{-0.04}^{+0.02}$ & 0.33$_{-0.1}^{+0.11}$ \\[5pt] 

 \hline
 \end{tabular} 	
		\label{tab.bcbhs4}
		\smallskip
		{}
	\end{minipage}
	\end{table*}

\begin{table*}
	\centering
		\begin{minipage}{140mm}
		\caption{Average BCs for pulsar X-ray binaries and intrinsic \nh{} in 5 Eddington ratio bins.}
		\begin{tabular}{@{}ccccccccc@{}}
			\hline 
			 	N& $\log_{min} \frac{L_{bol}}{L_{edd}}$ &$\log_{max} \frac{L_{bol}}{L_{edd}}$&S1 & H1& B1 & S2& H2 & B2 \\
			 \hline
1 &-3.54 & -2.79&0.06$_{-0.02}^{+0.04}$ & 0.57$\pm0.03$ & 0.62$_{-0.03}^{+0.05}$ & 0.44$_{-0.09}^{+0.04}$ & 0.25$_{-0.05}^{+0.04}$ & 0.7$_{-0.07}^{+0.04}$ \\[5pt] 
2 &-2.79 & -2.04&0.03$_{-0.01}^{+0.03}$ & 0.49$_{-0.06}^{+0.1}$ & 0.52$_{-0.06}^{+0.14}$ & 0.46$\pm0.05$ & 0.28$_{-0.06}^{+0.05}$ & 0.75$\pm0.05$ \\[5pt] 
3 &-2.04 & -1.29&0.02$\pm0.01$ & 0.42$_{-0.13}^{+0.05}$ & 0.45$_{-0.14}^{+0.05}$ & 0.4$_{-0.11}^{+0.05}$ & 0.32$_{-0.03}^{+0.02}$ & 0.73$_{-0.13}^{+0.04}$ \\[5pt] 
4 &-1.29 & -0.55&0.02$\pm0.01$ & 0.35$_{-0.1}^{+0.05}$ & 0.37$_{-0.1}^{+0.06}$ & 0.34$_{-0.07}^{+0.03}$ & 0.33$_{-0.05}^{+0.04}$ & 0.67$_{-0.08}^{+0.06}$ \\[5pt] 
5 &-0.55 & 0.2&0.02$_{-0.01}^{+0.02}$ & 0.32$\pm0.07$ & 0.33$_{-0.05}^{+0.1}$ & 0.32$_{-0.08}^{+0.13}$ & 0.32$_{-0.08}^{+0.1}$ & 0.58$_{-0.09}^{+0.28}$ \\
\hline
 \end{tabular} 	
		\label{tab.bcpulsars_in}
		\smallskip
		{}
	\end{minipage}
	\end{table*}

\begin{table*}
	\centering
		\begin{minipage}{140mm}
		\caption{Average BCs for pulsar X-ray binaries and fiducial \nh{} in 5 Eddington ratio bins. }
		\begin{tabular}{@{}ccccccccc@{}}
			\hline  
			 	N& $\log_{min} \frac{L_{bol}}{L_{edd}}$ &$\log_{max} \frac{L_{bol}}{L_{edd}}$ &S1 & H1& B1 & S2& H2 & B2 \\
			 \hline
1&-3.5 & -2.75&0.16$_{-0.05}^{+0.06}$ & 0.51$_{-0.04}^{+0.06}$ & 0.68$_{-0.06}^{+0.02}$ & 0.38$_{-0.07}^{+0.08}$ & 0.22$_{-0.04}^{+0.05}$ & 0.59$_{-0.08}^{+0.1}$ \\[5pt] 
2&-2.75 & -2.01&0.08$\pm0.03$ & 0.48$_{-0.07}^{+0.09}$ & 0.56$_{-0.08}^{+0.12}$ & 0.43$_{-0.05}^{+0.04}$ & 0.26$_{-0.05}^{+0.06}$ & 0.71$_{-0.07}^{+0.06}$ \\[5pt] 
3&-2.01 & -1.26&0.06$\pm0.02$ & 0.42$_{-0.13}^{+0.06}$ & 0.48$_{-0.15}^{+0.05}$ & 0.38$_{-0.09}^{+0.05}$ & 0.31$_{-0.03}^{+0.02}$ & 0.68$_{-0.1}^{+0.06}$ \\[5pt] 
4&-1.26 & -0.52&0.04$_{-0.02}^{+0.05}$ & 0.35$_{-0.1}^{+0.04}$ & 0.38$_{-0.12}^{+0.1}$ & 0.33$_{-0.06}^{+0.03}$ & 0.32$_{-0.06}^{+0.04}$ & 0.65$_{-0.09}^{+0.06}$ \\[5pt] 
5&-0.52 & 0.23&0.04$_{-0.03}^{+0.09}$ & 0.33$_{-0.07}^{+0.05}$ & 0.36$_{-0.05}^{+0.14}$ & 0.28$_{-0.05}^{+0.16}$ & 0.28$_{-0.07}^{+0.12}$ & 0.54$_{-0.08}^{+0.3}$ \\[5pt] 

\hline
 \end{tabular} 	
		\label{tab.bcpulsars}
		\smallskip
	\end{minipage}
	\end{table*}

\begin{table*}
	\centering
		\begin{minipage}{140mm}
		\caption{Average BCs for ULXps for bins 6 and 7 shown in left panel of Fig. \ref{fig.pulsars_ulxs} and fiducial \nh{}.}
		\begin{tabular}{@{}ccccccccc@{}}
			\hline  
			 	N& $\log_{min} \frac{L_{bol}}{L_{edd}}$ &$\log_{max} \frac{L_{bol}}{L_{edd}}$ &S1 & H1& B1 & S2& H2 & B2 \\
			 \hline
6 &0.23 & 1.48  	   &0.34 & 0.44 & 0.79 & 0.31 & 0.08 & 0.4 \\[5pt] 
7 &1.48 & 3.11  	   &0.16$_{-0.06}^{+0.08}$ & 0.58$_{-0.22}^{+0.07}$ & 0.85$_{-0.34}^{+0.01}$ & 0.37$_{-0.12}^{+0.08}$ & 0.08$_{-0.01}^{+0.0}$ & 0.44$_{-0.11}^{+0.09}$ \\[5pt] 
\hline
 \end{tabular} 	
		\label{tab.bculxps}
		\smallskip
	\end{minipage}
	\end{table*}

\begin{table*}
	\centering
		\begin{minipage}{140mm}
		\caption{Average BCs for ULXs assuming they are all NS for bins 6, 7, and 8 shown in the middle panel of Fig. \ref{fig.pulsars_ulxs} and fiducial \nh{}.}
		\begin{tabular}{@{}ccccccccc@{}}
			\hline  
			 	N& $\log_{min} \frac{L_{bol}}{L_{edd}}$ &$\log_{max} \frac{L_{bol}}{L_{edd}}$ &S1 & H1& B1 & S2& H2 & B2 \\
			 \hline
6 &0.23 & 1.66  	   &0.34 & 0.44 & 0.79 & 0.31 & 0.08 & 0.4 \\[5pt] 
7&1.66 & 1.87  	   &0.26$_{-0.1}^{+0.1}$ & 0.47$_{-0.05}^{+0.11}$ & 0.78$_{-0.08}^{+0.04}$ & 0.33$_{-0.08}^{+0.06}$ & 0.09$\pm0.03$ & 0.43$_{-0.11}^{+0.08}$ \\[5pt] 
8 &1.87 & 3.11  	   &0.22$_{-0.03}^{+0.06}$ & 0.51$_{-0.02}^{+0.14}$ & 0.75$_{-0.04}^{+0.12}$ & 0.37$_{-0.04}^{+0.05}$ & 0.1$_{-0.03}^{+0.05}$ & 0.49$_{-0.09}^{+0.03}$ \\[5pt] 
\hline
 \end{tabular} 	
		\label{tab.bculxns}
		\smallskip
	\end{minipage}
	\end{table*}

\begin{table*}
	\centering
		\begin{minipage}{140mm}
		\caption{Average BCs for ULXs assuming they are all BHs for bin 6 shown in the right panel of Fig. \ref{fig.pulsars_ulxs} and fiducial \nh{}.}
		\begin{tabular}{@{}ccccccccc@{}}
			\hline  
			 	N& $\log_{min} \frac{L_{bol}}{L_{edd}}$ &$\log_{max} \frac{L_{bol}}{L_{edd}}$ &S1 & H1& B1 & S2& H2 & B2 \\
			 \hline
6 &0.23 & 1.9    	   &0.24$_{-0.05}^{+0.07}$ & 0.5$_{-0.05}^{+0.11}$ & 0.76$_{-0.06}^{+0.07}$ & 0.35$\pm0.06$ & 0.11$_{-0.04}^{+0.03}$ & 0.47$_{-0.1}^{+0.05}$ \\[5pt]
\hline
 \end{tabular} 	
		\label{tab.bculxbh}
		\smallskip
	\end{minipage}
	\end{table*}

\begin{table*}

	\centering
		\begin{minipage}{140mm}
		\caption{Average optical V-band to X-ray 2-10 keV band ratios as a function of Eddington ratio $L_{(2-10\ keV)}/L_{edd}$ for our sample of BHXBs.}
		\begin{tabular}{@{}cccccc@{}}
			\hline  
			 N & $\log_{min} \frac{L2-10}{L_{edd}}$ &$\log_{max} \frac{L2-10}{L_{edd}}$ & $\frac{L_V}{L_X}$ (p=0.75) & $\frac{L_V}{L_X}$ (p=0.65) & $\frac{L_V}{L_X}$ (p=0.55) \\
			 &&&$\times10^{-5}$&$\times10^{-4}$&$\times10^{-3}$\\
			\hline
			
			1 & -4.52 & -4.1   &  4.05$_{-2.21}^{+5.88}$ & 4.05$_{-2.07}^{+5.2}$ & 9.32$_{-4.45}^{+8.64}$ \\[5pt]
2 & -4.1 & -3.67  & 2.41$_{-1.42}^{+6.16}$ & 2.36 $_{-1.26}^{+4.5}$ & 4.61$_{-2.15}^{+6.28}$ \\[5pt] 
3 & -3.67 & -3.24  & 3.4$_{-2.69}^{+12.28}$ & 2.87$_{-1.98}^{+8.96}$ & 5.64$_{-3.31}^{+21.66}$ \\[5pt]
4 & -3.24 & -2.81  & 0.92$_{-0.4}^{+4.68}$ & 1.03$_{-0.42}^{+3.85}$ & 2.78$_{-1.06}^{+4.67}$ \\[5pt] 
5 & -2.81 & -2.39  & 1.98$_{-1.19}^{+3.2}$ & 1.77$_{-0.87}^{+2.27}$ & 3.54$_{-1.21}^{+2.91}$ \\[5pt] 
6 & -2.39 & -1.96  & 12.36$_{-7.99}^{+9.81}$ & 10.58$_{-6.35}^{+10.28}$ & 20.96$_{-12.88}^{+25.76}$ \\[5pt]
7 & -1.96 & -1.53  & 3.44$_{-1.11}^{+6.94}$ & 4.28$_{-1.12}^{+6.33}$ & 14.5$_{-4.21}^{+11.33}$ \\[5pt]
8 & -1.53 & -1.1  & 2.3$_{-1.21}^{+0.88}$ & 2.96$_{-1.4}^{+1.1}$ & 11.22$_{-4.23}^{+3.63}$ \\[5pt] 	 
9 & -1.1 & -0.68  & 1.67$_{-0.77}^{+0.66}$ & 2.19$_{-0.83}^{+0.77}$ & 7.95$_{-2.14}^{+3.56}$ \\[5pt] 
10 &-0.68 & -0.25  & 0.96$_{-0.35}^{+0.47}$ & 1.45$_{-0.37}^{+0.62}$ & 6.49$_{-0.71}^{+2.47}$ \\[5pt]
\hline
 \end{tabular} 	
		\label{tab.fxfopt}
		\smallskip
		{}
	\end{minipage}
	\end{table*}


\bsp	
\label{lastpage}
\end{document}